\documentclass[
reprint,
showkeys,
aps,
prd,
longbibliography,
nofootinbib,
twocolumn
]{revtex4-2}
\usepackage[utf8]{inputenc}
\usepackage{amsmath}
\usepackage{amssymb}
\usepackage{xstring}
\usepackage{siunitx}
\usepackage{slashed}
\usepackage{dsfont}
\usepackage{upgreek}
\usepackage{bm}
\usepackage{mathtools}
\usepackage{layouts}

\usepackage[hyperref,dvipsnames]{xcolor}

\usepackage[
colorlinks,
pdfpagelabels,
breaklinks,
pdfstartview=FitH,
bookmarksopen=true,
bookmarksnumbered=true,
bookmarksopenlevel=2,
plainpages=false,
hypertexnames=false,
citecolor=blue,
linkcolor=blue,
urlcolor=blue,
pdftitle={Constraining fermionic condensates}
,pdfauthor={Gergely Endrodi, Kovacs Tamas, Gergely Marko, Laurin Pannullo}
]{hyperref}

\usepackage[T1]{fontenc} 
\usepackage{xparse}
\usepackage[capitalise]{cleveref}

\newcounter{numrefs}
\makeatletter
\newcommand{\Rcite}[1]{%
	\setcounter{numrefs}{0}
	\@for\@temp:=#1\do{\stepcounter{numrefs}}
	\ifnum\value{numrefs}>1%
	Refs.~\cite{#1}%
	\else%
	Ref.~\cite{#1}%
	\fi%
}
\makeatother

\usepackage{dsfont,slashed}
\usepackage{upgreek}
\usepackage{mathtools}

\NewDocumentCommand{\dr}{ o o } {%
	\mathop{}\!\mathrm{d}%
	\IfNoValueTF {#2} {%
	}
	{%
		^{#2}%
	}
	\IfNoValueTF {#1} {%
	}
	{%
		#1\,%
	}%
}

\makeatletter
\newcommand{\vast}{\bBigg@{4}}
\newcommand{\Vast}{\bBigg@{5}}
\makeatother

\newcommand{\D}{\mathcal{D}}

\newcommand{\dd}{\textmd{d}}
\DeclareMathOperator{\Tr}{Tr}

\newcommand{\eu}{\mathrm{e}}
\newcommand{\iu}{\mathrm{i}}

\newcommand{\expv}[1]{\left\langle#1\right\rangle}
\newcommand{\expConstrained}[2][\phi]{\left\langle #2 \right\rangle_{#1}}

\newcommand{\trone}{\mathcal{M}}
\newcommand{\trtwo}{\chi}
\newcommand{\trthree}{\kappa}

\newcommand{\HS}{\rho}   

\newcommand{\DDR}{\eta}      
\newcommand{\DDRExtreme}{ \bar{\DDR} }

\newcommand{\Sb}{S_{\rm B}}
\newcommand{\Sf}{S_{\rm F}}
\newcommand{\Z}{\mathcal{Z}}
\newcommand{\Zphi}{\mathcal{Z}_{\phi}}
\newcommand{\Zphiappr}{\mathcal{Z}_{\phi}^{\rm approx}}
\newcommand{\Zeta}{\widetilde\Z_\eta}

\newcommand{\ZGN}{\mathcal{Z}_{\rm \chi GN}}
\newcommand{\ZGNphi}{\mathcal{Z}_{\rm \chi GN\,\phi}}

\newcommand{\generatingFunction}{G}
\newcommand{\generatingFunctionApprox}{\generatingFunction^{\mathrm{approx}}}
\newcommand{\constraintDerivative}{j}

\newcommand{\phimin}{\bar\phi}

\newcommand{\UQeff}{\Gamma}

\newcommand{\U}{{{\cal U}}}

\newcommand{\uSmall}{{{ u}}}
\newcommand{\uSmallapprox}{\uSmall_{\mathrm{approx}}}

\newcommand{\m}{\Sigma} 
\newcommand{\Nm}{{N_{\m}}}
\newcommand{\Nmprod}[1][a]{\prod_{#1}}
\newcommand{\bilinearMatrix}{\sigma}
\newcommand{\source}{m}

\DeclareMathOperator{\Real}{Re}
\DeclareMathOperator{\Imag}{Im}

\newcommand{\coloneqq}{\equiv}

\newcommand{\Ns}{N_s}
\newcommand{\Nf}{N_f}
\newcommand{\vacScale}{\bar{\rho}}

\usepackage[acronym]{glossaries-extra}
\glssetcategoryattribute{acronym}{nohyperfirst}{true}
\setabbreviationstyle[acronym]{long-short}
\setabbreviationstyle[ignored]{long-noshort}

\newacronym{qcd}{QCD}{Quantum Chromodynamics}
\newacronym{gn}{GN}{Gross-Neveu}
\newacronym{chign}{$\chi$GN}{chiral Gross-Neveu}

\begin{document}

\title{ Constraining fermionic condensates}

\author{Gergely Endr\H{o}di$^{1,2}$}
\author{Tam\'as G.\ Kov\'acs$^{1,3}$}
\author{Gergely Mark\'o$^{2}$}
\author{Laurin Pannullo$^{2}$}

\email{gergely.endrodi@ttk.elte.hu}
\email{tamas.gyorgy.kovacs@ttk.elte.hu}
\email{gmarko@physik.uni-bielefeld.de}
\email{lpannullo@physik.uni-bielefeld.de}

\affiliation{$^1$Institute of Physics and Astronomy, ELTE E\"otv\"os Lor\'and University, P\'azm\'any P.\ s\'et\'any 1/A, H-1117 Budapest, Hungary}
\affiliation{$^2$Fakult\"at f\"ur Physik, Universit\"at Bielefeld, D-33615 Bielefeld, Germany}
\affiliation{$^3$Institute for Nuclear Research, Bem t\'er 18/c, H-4026 Debrecen, Hungary}

\begin{abstract}
    We study spontaneous symmetry breaking in quantum field theories with fermionic order parameters and construct, for the first time in the literature, the constraint effective potential for it. 
    The Grassmann-valued constraint we encounter is handled using its large-volume  expansion, corresponding to a saddle-point approximation.
    We test the method in the chiral Gross-Neveu model and demonstrate its consistency with the standard approach.
    The machinery we developed opens up a new avenue to investigate the spontaneous symmetry breaking and restoration in field theories, in particular for the chiral symmetry breaking in the strong interactions.
\end{abstract}

\maketitle
\flushbottom
	
\section{Introduction}
    Spontaneous symmetry breaking is ubiquitous in nature, appearing as an underlying non-perturbative concept in a multitude of physical systems ranging from condensed matter physics to elementary particle physics~\cite{Brauner:2010wm,Maas:2017wzi}. 
    The standard theoretical description of spontaneous breaking and the simulation of spontaneously broken theories in practice entails a double limit procedure: an extrapolation to the thermodynamic limit, followed by an extrapolation to vanishing explicit breakings in order to find the realized value $\phimin$ of the order parameter.
    Carrying out this program for quantum field theories with spontaneously broken continuous symmetries in particular, is computationally challenging due to the presence of massless Goldstone degrees of freedom. 
    
    A prime example for this type of behavior is \gls{qcd}, the theory of the strong interactions, in the limit that the masses of the light quarks vanish and chiral symmetry is exact.
    In this case, the order parameter is the chiral condensate, constructed from the fermionic fields, and to define it properly, the
    infinite volume limit must be taken prior to the chiral limit. 
    In fact, the nature of the finite temperature \gls{qcd} transition in the chiral limit is one of the few remaining open questions of zero-density \gls{qcd} that lattice simulations could not yet answer fully~\cite{Aarts:2023vsf}.
    A particularly challenging aspect in this context is the effective restoration of the anomalous axial symmetry and its effect on QCD thermodynamics~\cite{Aarts:2023vsf}.
    Even for the physical case, where quarks are massive, the chiral limit of \gls{qcd} imposes nontrivial bounds on the behavior of \gls{qcd} at nonzero baryon density~\cite{Hatta:2002sj}, i.e.\ the phase diagram of the theory and the anticipated critical endpoint on it, sought for in contemporary experimental campaigns~\cite{Du:2024wjm}.
    Further theories that feature spontaneous breaking, albeit via bosonic order parameters, include the electroweak theory and extensions of the Standard Model, e.g.\ in the Peccei-Quinn sector. 
    
    As an alternative to the computationally challenging double limit procedure, we propose to study fermionic theories with spontaneously broken continuous symmetries using the constraint effective potential $\Omega(\phi)$~\cite{Fukuda:1974ey,ORaifeartaigh:1986axd}. Then, the parameter characterizing the partition function is not the explicit breaking parameter but the value $\phi$ that the order parameter is constrained to.
    In this setup, the spontaneous symmetry breaking manifests itself in the flatness of the effective potential (in the infinite volume limit) on the disk $|\phi|<\phimin$.
    In finite volumes, in turn, the potential has local maxima at $|\phi|<\phimin$ and a continuum of minima (a ``valley'') at $|\phi|=\phimin$, reminiscent of the classical Mexican hat potential. 
    The advantage of our present approach is that $\phimin$ can be found by locating the edge of the disk, so that the twofold extrapolation procedure of the standard method is replaced by a single, infinite volume extrapolation, together with an interpolation in $\phi$.
    
    This method has been developed and tested for the case of scalar field theories, e.g.\ the $\mathrm{O}(2)$ model in three dimensions~\cite{Endrodi:2021kur}.
    Its generalization to fermionic condensates is highly non-trivial: it requires the definition of constraints with Grassmann-valued arguments as well as the development of novel simulation algorithms.
    In this paper we construct, for the first time, the constraint effective potential for fermionic order parameters in quantum field theories with continuous symmetries.
    We demonstrate our method in the \gls{chign} model, but our formalism is applicable in a wide range of other fermionic theories -- explicitly including \gls{qcd}.
    Previously, we applied the method to the \gls{gn} model, which has a discrete spontaneously broken symmetry, see \Rcite{Endrodi:2023est}.
    We note that \Rcite{Azcoiti:1995dq, Luo:2004wx} apply a similar constrained path integral approach to fermionic systems, but their approach differs significantly from ours in key steps.
    We will elaborate on the differences in \cref{sec:realisation}. 
    
    In the study of the constrained $\mathrm{O}(2)$ model in three dimensions~\cite{Endrodi:2021kur}, it was found that within the disk, the constrained path integral is dominated by inhomogeneous field configurations with their field values distributed along the valley.
    Similarly, we find that inhomogeneous fermionic condensates describe the flattening region in fermionic systems.
    We note that theses inhomogeneous condensates are distinct from exotic inhomogeneous condensates \cite{Brodsky:2009zd,Brodsky:2010xf,Brodsky:2012ku} that arise as a thermodynamical phase in strongly interacting matter, e.g., at nonzero quark density~\cite{Sadzikowski:2000ap,Thies:2003kk,Nakano:2004cd,Nickel:2009wj,Buballa:2014tba}, magnetic fields~\cite{Brauner:2016pko} or in the heavy-dense limit~\cite{Akerlund:2016myr}.
    
    This paper is structured as follows:
    In \cref{sec:ConstraintTheory}, we introduce the considered class of theories, the effective potential and the constraint potential.
    This is followed by \cref{sec:realisation}, where we discuss explicit realizations of the constrained path integral for fermionic order parameters.
    We introduce the  two-dimensional \gls{chign} model in \cref{sec:chiGN} and present numerical results of its constraint potential obtained via lattice field theory in \cref{sec:results}.
    We draw conclusions and give an outlook in \cref{sec:conclusion}.
    \cref{app:foolsGold} discusses formal aspects about different representations of $\delta$-distributions with Grassmann-valued arguments.

\section{The constraint effective potential as a probability distribution}
\label{sec:ConstraintTheory}

    We consider a $d$-dimensional quantum field theory with a continuous symmetry that is broken spontaneously by a fermionic order parameter.
    The partition function of the system is written in terms of the Euclidean path integral
    \begin{equation}
    \begin{split}
        \Z&=\int \D U\, \D\bar\psi \D\psi \, \eu^{-S_{\rm B}[U]+S_{\rm F}[\bar\psi,\psi,U]}\\
        &= \int \D U\,\eu^{-S_{\rm B}[U]} \det Q\,,
    \end{split}
    \label{eq:partfunc1}
    \end{equation}
    where $U$ denotes bosonic fields (i.e.\ gluons in \gls{qcd}) and $\bar\psi,\psi$ represent fermionic fields.
    The purely bosonic part $S_{\rm B}$ of the action is kept general, while we require the fermionic part $S_{\rm F}$ to be of the bilinear form
    \begin{equation}
    \label{eq:fermionAction}
        S_{\rm F}[\bar\psi,\psi,U] = \int \dr[x][d]
        \bar\psi(x) Q
        \psi(x)\,,
    \end{equation}
    which allowed us to perform the fermionic path integral in the second step of \cref{eq:partfunc1}.
    Here, $Q$ is a matrix in Dirac space that depends on the boson fields $U$ and which we will refer to as the Dirac operator; its particular form is not important at the moment.
    Furthermore, we only consider theories that satisfy $\det Q\in\mathds{R}^+$ i.e.\ that do not suffer from a complex action problem.
    
    The fermionic action exhibits a global continuous symmetry, characterized by the group $G$, which is broken spontaneously according to the pattern $G\to H\subset G$ (i.e.\ $\mathrm{SU}(2)_{\mathrm{L}} \times \mathrm{SU}(2)_{\mathrm{R}}\to\mathrm{SU}(2)_{\mathrm V}$ in the case of chiral two flavor \gls{qcd}).
    The order parameter for this symmetry breaking pattern is the general fermionic bilinear
    \begin{equation}
    \label{eq:orderPars}
         \m_a\coloneqq\frac{1}{V}\int \dr[x][d]
         \bar\psi(x) \bilinearMatrix_a\psi(x)\,,
    \end{equation}
    where $V$ denotes the space-time volume and the index $a$ runs over the $\Nm$ possible directions  of the order parameter in the internal (flavor and/or Dirac) space,
    specified by the matrices $\bilinearMatrix_{a}$.
    The latter are chosen so that $\m_a$ are Hermitean.
    
    The expectation value of the order parameter with respect to the path integral~\labelcref{eq:partfunc1}
    vanishes in any finite volume, since all directions of the order parameter are weighted symmetrically in $\Z$.
    The standard approach to define the expectation value is to include an explicit symmetry breaking source $\source$,
    \begin{align}
    \begin{split}
        \Z(\source)={}&
        \int \D U\D\bar\psi \D\psi\,\eu^{-S_{\rm B}[U]+S_{\rm F}[\bar\psi,\psi,U]+V\source_a\m_a}\\
        ={}&\int \D U \,\eu^{-S_{\rm B}[U]}\, \det(Q+m_a\sigma_a)
        \,,
        \label{eq:Zmdef}
    \end{split}
    \end{align}
    where in the second step the fermionic path integral was again carried out. 
    Expectation values with respect to $\Z(\source)$ are denoted below by $\langle\cdot\rangle$.
    The associated potential reads
    \begin{equation}
    \label{eq:WDefinition}
        W(\source) \equiv \frac{1}{V} \log \Z(\source)\,.    
    \end{equation}
    The physically realized value $\phimin$ of the order parameter is obtained by differentiating $W$ with respect to the source in the limit $\source\to0$ {\it after} the thermodynamic limit has been taken,
    \begin{equation}
        \phimin = \lim_{\source\to0}\lim_{V\to\infty} \frac{\partial W(\source)}{\partial \source_a}\,.
    \label{eq:standardapproach}
    \end{equation}
    In this way, one element of the vector space spanned by the basis vectors~\labelcref{eq:orderPars} is chosen explicitly and the symmetry of $\Z$ ensures that for~\labelcref{eq:standardapproach}, the direction of $\source$ in this space is insignificant.

\subsection{The effective potential}

    Instead of describing the spontaneous symmetry breaking in terms of the source $\source_a$, an equivalent description is provided by the quantum effective potential. 
    This object is parameterized by the expectation value $\expv{\m_a}=\phi_a$ of the order parameter. 
    The source and the order parameter expectation value are therefore conjugate quantities, similar to the chemical potential and the particle number in the grand canonical and canonical ensembles for a thermodynamic system.
    
    In particular, the quantum effective potential is given through the Legendre transform of $W(\source)$ as
    \begin{equation}
    \label{eq:EffectivePotentialDef}
        \UQeff(\phi) \coloneqq \sup_\source \left[-W(\source) + \source_a \phi_a\right]\,.
    \end{equation}
    As can be shown easily, for a given $\phi$, the supremum in this expression selects exactly the source value $\source_a$, in the presence of which $\expv{\m_a}=\phi_a$ holds. 
    Taking a derivative of $\UQeff$ yields the selected source value
    \begin{align}
    \label{eq:derivOfEffectivePotential}
        \frac{\partial \UQeff(\phi)}{\partial \phi_a} = m_a\,,
    \end{align} 
    and the physically realized value $\phimin$ may be found in the thermodynamic limit by locating the largest $|\phi|$ for which $m=0$.
    The set of $\phi_a$ values considered as arguments of $\Gamma$ are chosen such that one reproduces the symmetry breaking pattern of interest, i.e.\ $G\to H$.
    We note that the Legendre transform in the definition~\labelcref{eq:EffectivePotentialDef} is valid as long as $W(m)$ is convex and results in a manifestly convex function~\cite{Haymaker:1983xk}, see also Refs.~\cite{ORaifeartaigh:1986axd,Alexandre:2012ht}.

    We note that typically, the effective potential $\UQeff(\phi)$ is interpreted as the sum of all one-particle-irreducible diagrams and therefore the generating functional of the proper $n$-point functions at vanishing external momenta. 
    In this paper we explore the effective potential and related quantities from a different, probabilistic point of view.
    To that end, we need to revisit the path integral~\labelcref{eq:partfunc1} and classify the configurations in it according to the value that the order parameter $\m_a$ takes on them.
    This will lead us to the concept of the constraint effective potential.
    
\subsection{The constraint potential}
\label{sec:constrpot}

    The approach through \cref{eq:standardapproach} is the standard method for the determination of the expectation value of the order parameter. In particular, it has been followed in most studies to investigate chiral symmetry breaking in \gls{qcd} and to determine the associated order parameter, the chiral condensate~\cite{Aarts:2023vsf}. However, this approach is numerically extremely challenging, as it requires a series of simulations with different sources and large volumes. The smaller the source, the larger volumes are necessary in order for the would-be Goldstone modes (i.e.\ pions) to fit in the box.
    
    An alternative approach is given by the constraint effective potential $\Omega(\phi)$. It provides a probabilistic interpretation for the order parameter in a finite volume and, as we argue below, it also allows for a computationally simpler determination of $\expv{\m_a}$.
    The constraint effective potential is defined via a constrained path integral $\Zphi$.
    The latter includes a Dirac $\delta$, which allows only field configurations that have the desired value for the volume averaged order parameter~\cite{Fukuda:1974ey,ORaifeartaigh:1986axd}.
    Here, we extend the definition of $\Omega(\phi)$ given in \Rcite{ORaifeartaigh:1986axd} to order parameters built from fermionic fields.
    Formally, this constrained path integral reads
    \begin{align}
    \label{eq:defZphi}
        \Zphi\coloneqq{}& \int\D U\D\bar\psi \D\psi\, \,\eu^{-S_{\rm B}[U]+S_{\rm F}[\bar\psi,\psi,U]}
        \prod_a \delta(\phi_a - \m_a)\, ,
    \end{align}
    where the index $a$ runs over the $\Nm$ independent directions of the order parameter as introduced in \cref{eq:orderPars}.
    
    The constraint effective potential associated to $\Zphi$ is given by 
    \begin{align}
    \label{eq:defOmega}
        \Omega(\phi)={}&-\frac{1}{V}\log\Zphi\,.
    \end{align}
    Notice that $\int\prod_a \dd \phi_a\, \Zphi=\Z$ holds trivially, thus, $\Zphi/\Z$ is the probability density 
    of the order parameter.
    Analogous to thermodynamic ensembles, $\Omega(\phi)$ can be interpreted as a microcanonical potential, whose independent variable is not the expectation value of the extensive quantity but its value taken on individual states.
    In \Rcite{ORaifeartaigh:1986axd} it was shown that for bosonic order parameters the constraint potential $\Omega$ and the usual effective potential $\UQeff$ agree in the $V\to\infty$ limit.
    The formal considerations given there are based on properties of the Legendre transform and hold in the fermionic case as well, and hence $\Omega(\phi)=\UQeff(\phi)$ in the thermodynamic limit. 
    In \cref{sec:trueSaddlepoint}, we will prove this equality using an alternative argument as well.
    
    The probabilistic interpretation for $\Omega(\phi)$ provides an intuitive physical picture for the realization of spontaneous symmetry breaking in a finite volume.
    In the absence of the constraint, the most probable field configurations are spatially homogeneous with magnitude $|\m| = \phimin$.
    These correspond to the minima of the classical potential that are connected by the action of the symmetry group $G$ (e.g.~points along the valley of the Mexican hat potential).
    Since this is the most favorable field configuration, the constraint effective potential also has a minimum here: if the constraint variable is set to this particular value $\phi_a=\phimin$, the constrained path integral is dominated by the homogeneous $|\m|=\phimin$ field configurations.
    
    For $|\phi|>\phimin$, the minimizing configuration remains homogeneous and is equal to the constrained value.
    On the other hand, for $|\phi|<\phimin$, the fields vary along the minima of the classical potential in such a way that the constraint on the average can be satisfied by these spatially inhomogeneous field configurations.
    It is energetically more favorable for the field to stay along minima of the potential, even if it has a cost in terms of the ``kinetic energy''.   
    
    If $G$ is a discrete symmetry group, these inhomogeneities are given by domains that fill a preferred fraction of the total volume so that the averaged order parameter takes the value $\phi$~\cite{Weinberg:1987vp}.
    Varying $\phi_a$ along one spatial axis from $+\phimin$ to $-\phimin$ thus continuously connects two homogeneous vacua via inhomogeneous ones and is tantamount to realizing a first-order phase transition between these two vacua via bubble formation.
    For a continuous symmetry group $G$, the orientation of the order parameter can change continuously, and thus the inhomogeneous configurations are spin-wave like deformations with a wavelength of the order of the linear size of the system.
    This was shown in the $\mathrm{O}(2)$ model in \Rcite{Endrodi:2021kur}, and we will demonstrate explicitly in \cref{sec:chiGN_configs} that the same phenomenon occurs for the case of fermionic order parameters.
    
    In both cases, the action cost of domain boundaries (discrete case) or order parameter gradients (continuous case) increases $\Omega(\phi)$ for $|\phi|<\phimin$ relative to its value along the valley at $|\phi|=\phimin$, giving  $\Omega(\phi)$ the form of a generalized Mexican hat potential.
    The valley location can be identified already in finite volumes by determining the values of $\phi$ for which the derivative of $\Omega$,
    \begin{align}
    \label{eq:derivOfConstraintPotential}
        \constraintDerivative_a \equiv \frac{\partial \Omega(\phi)}{\partial \phi_a}\, ,
    \end{align}
    vanishes.
    This derivative can be thought of as the expectation value of the source term in a similar way as in \cref{eq:derivOfEffectivePotential}.
    This way of defining the order parameter is simpler than through the intricate double limit involved in \cref{eq:standardapproach}, as has already been demonstrated in~\Rcite{Endrodi:2021kur} for scalar fields.
    
    In the infinite volume limit, the excess action associated with domain boundaries or with such order parameter gradients is suppressed compared to the non-kinetic (the potential) part of the action, thus $\Omega(\phi)$ becomes flat for $|\phi|<\phimin$ in the limit $V\to\infty$.
    Therefore, while in finite volumes $\Omega$ is not necessarily convex, it becomes convex (and, as discussed above, equal to $\UQeff$) in the thermodynamic limit.
    
    In summary, the general structure of the constraint effective potential for bosonic and fermionic order parameters is highly analogous.
    There is, however, a crucial difference between them that arises from the Grassmannian nature of the fermion fields and renders the discussion in the present case more complicated.
    Namely, in a finite volume (assuming a finite regularization or discretization), only a finite number of Grassmann numbers $\psi_n$ and $\bar\psi_n$ exist, with $n=1\ldots N$.
    Thus, the order parameter~\labelcref{eq:orderPars} is not a classical number: even though it commutes with any Grassmann number, it is nilpotent as $\m_a^{N+1}=0$. 
    The formal meaning of a $\delta$ distribution for arguments containing Grassmann numbers is well defined in the mathematical domain of superanalysis and can be found, e.g., in \Rcite{DeWitt:2012mdz}.
    Using this formal meaning, the fermionic path integral gives rise to a result which can only be understood as a distribution of $\phi$ and the interpretation in terms of the constraint potential is lost. However, an equivalent description in terms a Fourier transform can also be given.
    Some details of the superanalysis results are recalled in \cref{app:foolsGold}, as well as the starting point of \cref{sec:realisation} is derived.

\section{Realization of the constraint potential for fermionic order parameters\label{sec:realisation}}

    In this section we discuss specific, approximate realizations of the constraint potential. We will introduce an approximation scheme, which we prove to become exact in the infinite volume limit, but also make contact with other techniques previously used in the literature.
    
    In order to treat the Grassmann integral in \cref{eq:defZphi} analytically, we represent the Dirac $\delta$ via its Fourier transform,
    \begin{equation}
        \delta(\phi_a-\m_a) = \lim_{\varepsilon\to0}\int_{-\infty}^\infty \frac{\dr[\DDR_a]}{2\uppi}\,\eu^{i\DDR_a(\phi_a-\m_a)-\DDR_a^2\varepsilon/(2V)}\,,
    \label{eq:diracdeps}
    \end{equation}
    with no summation over $a$ in the exponent on the right hand side. In \cref{app:foolsGold} we show that such a representation behaves as expected in terms of the Grassmannian nature of $\m_a$. The real parameter $\varepsilon>0$, is included in order to ensure the convergence of the $\eta_a$ integral.\footnote{Note that including $\varepsilon>0$ in the Fourier transform on the right hand side of~\labelcref{eq:diracdeps} is equivalent to considering a nascent $\delta$ of the Gaussian form with width $\varepsilon/V$.} It is sent to zero at the end of the calculation and is normalized by the volume for later convenience.
    Inserting this representation in \cref{eq:defZphi} and exchanging the path integral with the integral over $\DDR$ and the $\varepsilon\to0$ limit yields the following expression for the constrained path integral
    \begin{equation}
        \label{eq:defZeta}
        \Zphi=\lim_{\varepsilon\to0}\int_{-\infty}^\infty \Nmprod[a] \frac{\dr[\DDR_a]}{2\uppi}\,\eu^{\iu\DDR_a\phi_a-\DDR_a^2\varepsilon/(2V)}\, \Zeta\,, 
    \end{equation}
    where 
    \begin{equation}
        \Zeta=\int \D U \,\D\bar\psi \D\psi \Nmprod[a]\,\eu^{-\iu\DDR_a \m_a}\,\eu^{-\Sb[U]+\Sf[\bar\psi,\psi,U]}\,,
    \end{equation}
    can be thought of as the characteristic function of $\Zphi$ as a probability distribution. Notice that $\Zeta=\Z(\source=-\iu\DDR/V)$ equals the unconstrained partition function~\labelcref{eq:Zmdef} in the presence of an imaginary explicit breaking source, pointing in a general direction in the order parameter space.
    
    The characteristic function $\Zeta$ is itself of interest, as it can be related to physical observables and evaluated directly, e.g., via the eigenvalues of the Dirac operator.
    This approach was followed in \Rcite{Azcoiti:1995dq} for the two-dimensional Schwinger model and for QCD on very coarse lattices in \Rcite{Luo:2004wx}.
    In both cases, the general form of $\Zeta$ was derived and compared to numerical results obtained by the full diagonalization of the Dirac operator.
    A subtle but important difference in the procedure used in these studies compared to our description is that \Rcite{Azcoiti:1995dq,Luo:2004wx} only introduced a single Dirac $\delta$ to enforce the constraint.
    This can be seen as an averaged version of our definition of $\Zphi$ over the other directions in the space of order parameters, affecting the probabilistic interpretation described in \cref{sec:constrpot}.
    Moreover, a full diagonalization of the Dirac operator in higher dimensions or on larger lattices is unfeasible.
    Instead, here we follow a different procedure and proceed by carrying out the integral over $\DDR$ approximately.

\subsection{Saddle-point approximation and Legendre transform}
\label{sec:trueSaddlepoint}
    
    We can recast the integral over $\DDR$ in \cref{eq:defZeta} to the form,
    \begin{align}
        \label{eq:etaIntegralI}
        \Zphi&=\int_{-\infty}^\infty \Nmprod[a] \frac{\dr[\DDR_a]}{2\uppi}\,\eu^{-V \U(\DDR)}\,, \\ \U(\DDR)&=-\frac{1}{V}\left(\iu\DDR_a\phi_a -\frac{\DDR_a^2\,\varepsilon}{2V} + \log \Zeta\right)\, . 
    \end{align}
    A common technique to calculate integrals of this type
    is the so-called saddle-point approximation \cite{Bender:1999box}, which becomes exact as $V\to \infty$. The method assumes that $\Real\, \U$ has a global minimum at $\DDR=\DDRExtreme$ and utilizes the fact that for large $V$, the integral is dominated by the contribution in the vicinity of this point, as the rest of the integrand is exponentially suppressed. 
    Expanding $\U$ around $\DDRExtreme$ facilitates the analytic evaluation of the integral.
    
    The point $\DDRExtreme$ is characterized by the $\Nm$ conditions of vanishing gradients 
    \begin{align}
    \label{eq:LaplaceConditionI}
        \left.\frac{\partial \,\U }{\partial \DDR_a}\right|_{\DDR=\DDRExtreme} = \frac{\iu}{V} \left[ \langle\trone_a(\DDRExtreme)\rangle_{\DDRExtreme} - \phi_a-\frac{\iu\DDRExtreme_a \varepsilon}{V}\right]= 0\,,
    \end{align}
    where
    \begin{align}
        \trone_a(\DDR)\equiv \frac{1}{V}\,\Tr \left[\left(Q- \frac{\iu\DDR_c}{V}\bilinearMatrix_{c}\right)^{-1}\!\!\bilinearMatrix_a\right]\,,
        \label{eq:tronedef}
    \end{align}
    and the expectation value $\langle\cdot\rangle_{\DDR}$ is taken with respect to $\Zeta$.
    Moreover, we assume that at the minimum all the eigenvalues of the Hessian $H_{ab}$ of $\U$
    have positive real parts (we get back to this point below).
    This Hessian reads
    \begin{align}
        H_{ab}(\DDR) = \frac{1}{V^2}\Big[&\langle \trtwo_{ab} (\DDR) \rangle_\DDR +\varepsilon \delta_{ab} + \langle \trone_a(\DDR)\trone_b(\DDR) \rangle_\DDR \nonumber \\  &-\langle \trone_a(\DDR) \rangle_\DDR \langle \trone_b(\DDR) \rangle_\DDR \Big]\,,
    \end{align}
    where we defined
    \begin{align}
        \trtwo_{ab}(\DDR)\equiv \frac{-1}{V}\Tr\left[\left(Q- \frac{\iu\DDR_c}{V}\bilinearMatrix_{c}\right)^{-1}\!\!\bilinearMatrix_a\left(Q- \frac{\iu\DDR_d}{V}\bilinearMatrix_{d}\right)^{-1}\!\!\bilinearMatrix_b\right]\, .
        \label{eq:trtwodef}
    \end{align}
    
    Notice that in the $\varepsilon\to0$ limit, \cref{eq:LaplaceConditionI} is reminiscent of the implicit condition used in the definition of the Legendre transform in \cref{eq:EffectivePotentialDef}, but for a general complex source $\source_a=-\iu\DDR_a/V$.
    Thus, the definiteness of $\UQeff(\phi)$ implies the existence of a solution of \cref{eq:LaplaceConditionI} at a purely imaginary $\DDRExtreme$, in the limit of vanishing $\varepsilon$.
    Expanding $\U$ up to second order in $\DDR$ around the expansion point $\DDRExtreme=\DDRExtreme(\phi)$ and carrying out the integral yields the saddle-point approximation,
    \begin{align}
    \label{eq:ZphiDDRLegendre}
        \Zphi \approx  \frac{ \eu^{-V \U(\DDRExtreme)}}{ \sqrt{ \det (2 \uppi V H(\DDRExtreme))} } \,,
    \end{align}
    which becomes exact in the infinite volume limit.
    As argued above, the minimum of $\U$ corresponds to the $\source$ appearing in the definition~\labelcref{eq:EffectivePotentialDef} of $\UQeff$, thus it follows that for any volume
    \begin{equation}
        \lim_{\varepsilon\to0}\, \U(\DDRExtreme(\phi))=\UQeff(\phi)\,.
    \end{equation}
    Owing to the saddle-point approximation becoming exact in the infinite volume limit, this construction ensures that
    \begin{equation}
        \Omega(\phi)=-\frac{1}{V} \log \Zphi
        \xrightarrow{V\to\infty}
        \UQeff(\phi)\,,
    \end{equation}
    relating the constraint potential to the quantum effective potential.
    This constitutes an equivalent alternative proof that $\Gamma=\Omega$ holds in the thermodynamic limit.
    Both here and in the original work~\cite{ORaifeartaigh:1986axd}, the core element of the proof is a saddle-point approximation for large volumes or, equivalently, a Legendre transform.\footnote{\label{fn:artefact}There is a delicate issue regarding the uniqueness of the solution to \cref{eq:LaplaceConditionI}.
    In the presence of an ultraviolet regulator $\Lambda$, this uniqueness is lost as solutions at $|\DDRExtreme|\gtrsim\Lambda$ emerge. This ambiguity arises analogously in the Legendre transform~\labelcref{eq:EffectivePotentialDef}, because $W(m)$ is not convex in the region $m\gtrsim\Lambda$.
    The proper definition of both $\Gamma$ and $\Omega$ is understood to involve only the solutions that do not decouple for $\Lambda\to\infty$.}
    
    In summary, we saw that a construction reproducing the effective potential can be given in terms of the constraint potential.
    Our goal, however, is to derive an expression that can be treated in simulations already in finite volumes.
    In the next subsection we show how the formulation in terms of the saddle-point approximation can be used to construct exactly that.

\subsection{Saddle-point approximation and large volume expansion}
\label{sec:ApproximateSaddlepointApproximation}
    
    We have shown in \cref{sec:trueSaddlepoint} that the saddle-point approximation of the integral over $\DDR$ yields the quantum effective potential in the $V\to\infty$ limit.
    To obtain a quantity that is useful in actual calculations, we modify the described procedure.
    
    First, we carry out only the fermionic path integral in \cref{eq:defZeta} to obtain 
    \begin{align}
        \label{eq:etaIntegralExchanged}
        \Zphi&=\lim_{\varepsilon\to0}\int \D U\, \eu^{-\Sb[U]}\int_{-\infty}^\infty \Nmprod[a] \frac{\dr[\DDR_a]}{2\uppi} \exp\left[-V\uSmall(\DDR)\right] \,,\\ \uSmall(\DDR)&=\frac{-1}{V}\left[ \iu\DDR_a\phi_a -\frac{\DDR_a^2\varepsilon}{2V}
        + \log\det\left(Q- \frac{\iu\DDR_a}{V}\bilinearMatrix_{a}\right)\right]\, ,
    \end{align}
    so that we have a representation of the constraint on every bosonic configuration individually.
    
    Second, to study spontaneous symmetry breaking, we are ultimately interested in the flat region (disk) of $\UQeff$ and the location of its edges.
    In the infinite volume limit, this region corresponds to a Legendre transform with $\source=0$ (see \cref{eq:derivOfEffectivePotential}).
    
    Therefore, we choose to expand $\uSmall$ not around the true saddle-point, as done in the previous section, but rather around $\DDRExtreme=0$ as this is the correct expansion point in the infinite volume as well as $\varepsilon\to0$ limits for values of $\phi$ that correspond to the disk.
    The quadratic expansion of $\uSmall$ around $\DDR=0$ takes the form
    \begin{align}
    \uSmall(\DDR)={}&\uSmallapprox(\DDR)+\mathcal{O}(\DDR^3/V^3),\\
    \begin{split}
        \uSmallapprox(\DDR) ={}& -\frac{1}{V}\log\det Q +\frac{\iu \DDR_a }{V}  (\trone_a-\phi_a) \\
        &+ \frac{1}{2V^2} \,\DDR_a \left(\trtwo_{ab}+\varepsilon\delta_{ab}\right) \,\DDR_b\,,
        \label{eq:uSmallexpansion}
    \end{split}
    \end{align}
    where $\trone \equiv \trone(\eta=0)$ and $\trtwo \equiv \trtwo(\eta=0)$ from \cref{eq:tronedef,eq:trtwodef}.
    The expansion~\labelcref{eq:uSmallexpansion} inherently corresponds to an expansion in $1/V$, highlighting its importance for the thermodynamic limit.
    Replacing $\uSmall$ by $\uSmallapprox$ in \cref{eq:etaIntegralExchanged} yields the approximated constrained path integral,
    \begin{widetext}
    \begin{align}
        \Zphiappr={}&\lim_{\varepsilon\to0}\int \D U\, \eu^{-\Sb[U]}\int_{-\infty}^\infty \Nmprod[a] \frac{\dr[\DDR_a]}{2\uppi} \exp\left[-V\uSmallapprox(\DDR)\right] \nonumber\\
        ={}& \lim_{\varepsilon\to0}\int \D U\, \eu^{-\Sb[U]} \frac{\det Q}{\sqrt{\det (2\uppi(\trtwo+\varepsilon)/V)}} \exp\left[ -\frac{V}{2}(\phi_a-\trone_a) \left(\trtwo+\varepsilon\right)_{ab}^{-1} (\phi_b-\trone_b) \right]\,, \label{eq:etaIntegralBeforeII}
    \end{align}
    \end{widetext}
    where the Hessian simplified to $H=(\chi+\varepsilon)/V^2$.
    Along the contour of real $\DDR$, the chosen expansion point $\DDRExtreme=0$ is the minimum of $\uSmall$ and thus the largest contribution to the integral.\footnote{We will show in numerical examples in \cref{sec:resSaddlePoint}, that this holds up to discretization artifacts that arise for large values of $|\DDR|$, cf.\ footnote~\ref{fn:artefact}.}
    We stress that $\uSmall$ has a non-zero, purely imaginary derivative at $\DDR=0$, so that the present approach represents an approximation of the saddle-point approximation, based on an expansion point that is slightly away from the true saddle point in a finite volume. The limit $\varepsilon\to0$ finally can be safely taken in the formula \cref{eq:etaIntegralBeforeII} since $\trtwo$ itself is manifestly positive definite\footnote{Up to discretization artifacts that might make $\trtwo$ non-positive definite at finite values of the ultraviolet regulator.}
    to obtain
    \begin{widetext}
    \begin{align}
        \Zphiappr=\int \D U\, \eu^{-\Sb[U]} \frac{\det Q}{\sqrt{\det \left(2\uppi\trtwo/V\right)}} \exp\left[ -\frac{V}{2}(\phi_a-\trone_a)\trtwo_{ab}^{-1} (\phi_b-\trone_b) \right]\,.    \label{eq:etaIntegralII}
    \end{align}
    \end{widetext}
    
    A crucial difference between \cref{eq:etaIntegralII} and \cref{eq:ZphiDDRLegendre} is that the integral over the bosonic fields is not yet carried out.
    Thus, we formulated a representation of the constrained path integral in terms of a standard path integral with a modified action in terms of the bosonic fields (recall that $\trone$, $\trtwo$ and $Q$ are functionals of $U$), which naturally lends itself to established methods such as Monte-Carlo simulations.
    One sees that in \cref{eq:etaIntegralII}, the modified action reduces the weight of configurations for which $\trone$ deviates from the constraint value $\phi$.
    This expression constitutes our final result for the constrained fermionic partition function.
    Notice that $\int \prod_a \dd \phi_a \,\Zphiappr = \Z$ holds, thus $\Zphiappr/\Z$ is also a normalized probability distribution and retains the full information of the unconstrained path integral.
    
    \subsection{The approximated constrained path integral}
    \label{sec:approxprobdist}
    
    In this section we illustrate certain properties and aspects of the approximated constrained path integral \labelcref{eq:etaIntegralII} that we derived in \cref{sec:ApproximateSaddlepointApproximation}. 

\subsubsection{Comparison of the exact and approximated constrained path integral via moments\label{sec:moments}}

    The purpose of this section is to quantify the impact of the approximations made in \cref{sec:ApproximateSaddlepointApproximation} and to check whether one indeed recovers the exact distribution $\Zphi/\Z$ from $\Zphiappr/\Z$ in the infinite volume limit.
    To do so, we consider both objects as probability distributions and compare their moments as a way to quantify differences between them.
    
    We define the moments of the probability distribution $\Zphi/\Z$ as \
    \begin{align}
        \label{eq:phiMoment}
        \mathfrak{m}_n = 
        \frac{1}{\Z}\int \prod_a \dd\phi_a \mathcal{Q}_n(\phi)\, \Zphi\,,
    \end{align}
    where
    \begin{align}
        \label{eq:calQDef}
        {\mathcal{Q}_n}(\phi) = \prod_b \phi_b^{n_b}
    \end{align}
    is a monomial built from different powers of the components of the vector $\phi$.
    Alternatively, these moments can be obtained by taking derivatives with respect to sources $\source$ of a generating function
    \begin{align}
        \generatingFunction(\source)=\frac{1}{\Z}\int \prod_a \dd\phi_a\,\eu^{V\source_a\phi_a}\Zphi
    \end{align}
    such that
    \begin{align}
        \mathfrak{m}_n = \left[\prod_a \frac{1}{V^{n_a}} \frac{\partial^{n_a}}{\partial\source_a^{n_a}} \right] \generatingFunction(\source)\Bigg|_{\source=0}.
    \end{align}
    
    Using these definitions with the exact constraint distribution $\Zphi/\Z$ from \cref{eq:defZphi}, we find that the generating function is given by the massive path integral
    $\generatingFunction(m) = \Z(m)/\Z$ and the moments trivially evaluate to
    \begin{equation}
    \label{eq:momentsFull}
       \mathfrak{m}_n=\expv{\mathcal{Q}_n(\m)}\,.
    \end{equation}
    
    To discuss the moments of the distribution $\Zphiappr/\Z$, we consider its generating function
    \begin{widetext}
    \begin{align}
        \generatingFunctionApprox(\source)=&{}\frac{1}{\Z}\int \prod_a \dd\phi_a\,\eu^{V\source_a\phi_a} \Zphi^{\rm approx}\nonumber\\
        ={}& \frac{1}{\Z}\int \prod_a \dd\phi_a \int \D U\,\eu^{-S_{\rm B}[U]}\frac{  \det Q}{\sqrt{\det (2\uppi\trtwo/V)}} \exp{\left[ V\source_a\phi_a-\frac{V}{2}(\phi_a-\trone_a) \trtwo^{-1}_{ab} (\phi_b-\trone_b) \right]}\nonumber\\
        ={}&\frac{1}{\Z}\int \D U\,\eu^{-S_{\rm B}[U]}\det Q\, \exp\left[V\source_a\trone_a+\frac{V}{2}\source_a \trtwo_{ab}\source_b\right]\nonumber\\
        ={}&\frac{1}{\Z}\int \D U\,\eu^{-S_{\rm B}[U]}\exp\left[\log\det\left(Q+\source_a\bilinearMatrix_a\right)+{\cal O}(\source^3)\right],
    \end{align}
    \end{widetext}
    where we identified the logarithm of the massive fermionic determinant up to $\mathcal{O}(m^3)$ corrections, by which this expression differs from the exact generating function $\Z(m)/\Z$.
    From this result, we expect moments $\mathfrak{m}_n$ with $n_a\leq2$ to be identical between the two distributions, while higher moments to be subject to corrections.
    
    To explicitly illustrate these corrections, we compare the first three moments for both the exact and the approximated constrained path integral of a single constraint $\delta(\phi_0-\m_0)$ in \cref{tab:momentComparison}.\footnote{One can certainly formulate these moments for an arbitrary number of constraints, but the expressions are rather involved and would unnecessarily complicate this simple comparison.}
    Corrections appear for all contributions that contain $Q^{-n}$ with $n>2$ as these are not present in $\generatingFunctionApprox$ and thus are not generated.
    These corrections, however, are such that they vanish in the limit $V\to\infty$ and thus one recovers all moments of the exact distribution.
    
    \begin{table*}[]
        \centering
        \begin{tabular}{c|l|l}
             $n$ & $\mathfrak{m}_n$ & $\mathfrak{m}_n-\mathfrak{m}_n^{\mathrm{approx}}$\\[3mm]\hline
             1 & $\expv{\m_0}\, = \frac{1}{V} \ \, \expv{\bar\psi\psi} \ \ \ \ \: = \frac{1}{V}\ \ \ \expv{\Tr Q^{-1} \vphantom{\left(\Tr Q^{-1}\right)^2} }$ & $0$ \\[3mm]
             2 & $\expv{\m_0^2}= \frac{1}{V^2}\expv{(\bar\psi\psi)^2} = \frac{1}{V^2} \left[\expv{\left(\Tr Q^{-1}\right)^2}-  \expv{\Tr Q^{-2}} \right]$ & $0$ \\[3mm]
             3 & $\expv{\m_0^3}= \frac{1}{V^3}\expv{(\bar\psi\psi)^3} =\frac{1}{V^3}\left[\expv{\left(\Tr Q^{-1}\right)^3}  - 3 \expv{\Tr Q^{-2} \Tr Q^{-1}} + 2 \expv{\Tr Q^{-3}}\right]$ & $\frac{2}{V^3} \expv{\Tr Q^{-3}}$
        \end{tabular}
        \caption{The first three moments of the distribution $\Zphi/\Z$ and their differences to the moments of the  approximated distribution $\Zphiappr/\Z$ for a constrained path integral with a single constraint $\delta(\phi_0-\m_0)$.}
        \label{tab:momentComparison}
    \end{table*}
    
    \subsubsection{Exactness of the constraint}
    Another way to gauge the impact of the approximation is to study the exactness of the constraint, i.e., how well the fermionic condensates $\expConstrained{\m}$ satisfy the constraint.\footnote{Note that $\expConstrained{\m}$ is evaluated according to a constrained path integral and as such is different from $\expv{\m}$ appearing in Sec~\ref{sec:moments}.}
    
    In an unconstrained path integral, the condensates are obtained by taking derivatives with respect to the specific sources as shown in \cref{eq:standardapproach}.
    Taking this approach with an exact constrained path integral, using $Q+\source_a\bilinearMatrix_a$ in place of $Q$ yields the expected result
    \begin{align}
    \label{eq:exactConstraintCondensate}
        \expConstrained{\m_a} = \frac{1}{V} \frac{\partial \log \Zphi(m)}{\partial \source_a}\Bigg|_{\source=0} = \phi_a\, ,
    \end{align}
    which holds for any volume.
    The expectation value $\expConstrained{\cdot}$ denotes an expectation value with respect to a constrained path integral.
    However, if the constrained path integral is represented by the approximated $\Zphiappr$, one finds the condensate expectation values
    \begin{align}
        \expConstrained{\m_a} ={}& \frac{1}{V} \frac{\partial \log \Zphiappr(m)}{\partial \source_a} \Bigg|_{\source=0}\nonumber\\
        ={}& \phi_a + \expConstrained{\left(\phi_b-\trone_b\right) \trtwo^{-1}_{bc}\,\trthree_{acd}
        \,  \trtwo^{-1}_{de} \left(\phi_e-\trone_e\right) }\nonumber\\
        &- \frac{1}{2V} \expConstrained{ \trtwo^{-1}_{cb}\trthree_{abc}}\, ,
    \label{eq:apprConstraintCondensate}
    \end{align}
    where $\trone$ and $\trtwo$ are given by \cref{eq:tronedef,eq:trtwodef} at $\DDR=0$ and 
    \begin{align}
        \trthree_{abc}=\frac{1}{V}\Tr\left[Q^{-1}\, \bilinearMatrix_a\, Q^{-1}\, \left(\bilinearMatrix_b\, Q^{-1}\, \bilinearMatrix_c+\bilinearMatrix_c Q^{-1}\bilinearMatrix_b\right)\right]\,.
    \end{align}
    It is a priori not obvious how large the additional terms in \cref{eq:apprConstraintCondensate} are compared to \cref{eq:exactConstraintCondensate} and whether they are suppressed in the infinite volume limit.
    We present numerical results on these extra contributions within a test case in \cref{sec:resCondensate} to show that the deviation $\expv{\m_a}_\phi-\phi_a$ is strongly suppressed for all values of $\phi$ that we are interested in. 
    
    \subsubsection{Derivatives of the constraint potential}
    The derivative of the quantum effective potential connects the expectation value $\phi_a$ and selected source term $\source_a$, see \cref{eq:derivOfEffectivePotential}.
    Similarly, we can explicitly compute the derivative~\labelcref{eq:derivOfConstraintPotential} of the constraint potential using the approximated constrained path integral~\labelcref{eq:etaIntegralII}, in terms of constrained expectation values,
    \begin{align}
        \constraintDerivative_a = \expConstrained{\trtwo^{-1}_{ab}(\phi_b-\trone_b)}\, .
    \end{align}
    This shows that a nonzero slope of the effective potential in $\phi$ arises from the deviation of the $\trone$ observables from the constraint values $\phi$.
    In the thermodynamic limit, the effective potential is flat i.e.\ its slope vanishes: therefore, for $V\to\infty$ either $\trone_b-\phi_b$ approaches zero or it becomes an eigenvector of $\trtwo^{-1}$ with zero eigenvalue.
    
    These derivatives can be used to reconstruct the constraint potential $\Omega$ that might not be directly accessible in some calculations such as Monte Carlo simulations \cite{Endrodi:2021kur}.

\subsection{Comparison to a constraint of \boldmath$\trone_a$ or density of states}

    In unconstrained calculations, one finds that 
    \begin{align}
        \expv{\m_a} = \expv{\trone_a} \,,
    \end{align}
    and therefore $\trone_a$ is often considered to be equivalent to the order parameter.
    As such, one might be inclined not to constrain $\m_a$, but instead introduce constraints of the form
    $\delta(\phi_a - \trone_a)$ outside the fermionic, but inside the bosonic path integral.
    Considering this kind of constrained path integral, representing the Dirac $\delta$ again via Gaussian forms,
    \begin{align}
    \label{eq:troneConstraintNascent}
        \prod_a \delta(\phi_a - \trone_a) = \lim_{\varepsilon\to0}\,\prod_a\frac{\exp\left[-\frac{V(\phi_a-\trone_a)^2}{2\varepsilon}\right]}
    {\sqrt{2 \uppi \,\varepsilon/V}}\,,
    \end{align}
    after integrating over the fermionic degrees of freedom, yields a constrained path integral
    \begin{align}
    \label{eq:ZphiDOS}
        \Zphi = \lim_{\varepsilon\to0}\int \D U\, \eu^{-\Sb[U]}\det Q \prod_a \frac{\exp\left[-\frac{V(\phi_a-\trone_a)^2}{2\varepsilon}\right]}{\sqrt{2 \uppi \,\varepsilon/V}}\,.
    \end{align}
    
    Clearly, this representation is equivalent to \cref{eq:etaIntegralII}, upon replacing the susceptibility matrix $\chi_{ab}$ (a functional of the bosonic fields) by $\varepsilon \delta_{ab}$ (a constant matrix).
    This form practically amounts to binning the $\trone_a$ observable with a resolution of $\varepsilon$ and is therefore akin to the density of states formulation \cite{Langfeld:2016kty}. The latter is commonly used in, e.g., investigations of Yang-Mills theory \cite{Lucini:2023irm}
    , but has also been employed for observables similar to $\trone_a$ in \Rcite{Endrodi:2018zda}.
    The absence of $\chi$ in \cref{eq:ZphiDOS} already indicates that fluctuations of the fermionic condensate $\m_a$ are not taken into account properly.
    This becomes evident when one repeats the analysis of \cref{sec:approxprobdist} of the moments of the resulting probability distribution.
    These are simply given by 
    \begin{align}
    \label{eq:momentsTrOne}
        \mathfrak{m}_n=\begin{cases}
            \expv{\trone_a}& \text{ for } n_b=\delta_{ba},\\
            \expv{\mathcal{Q}_n(\trone)}+{\cal O}(\varepsilon/V)&\text{ for }|n|>1\,.
        \end{cases}
    \end{align}
    Comparing to \cref{eq:momentsFull}, we find that the first moments with $|n|=1$ agree, but higher moments -- quantifying the fluctuations of $\m$ -- differ.
    By comparison of the moments, we recognize that the probability distribution that is obtained via the formulation of the constrained path integral~\labelcref{eq:etaIntegralII} carries more information in a finite volume than the density of states approach.
    This discrepancy remains in the infinite volume limit for the case of a spontaneously broken symmetry, where the distribution $\Zphi/\Z$ has a nonzero variance even for $V\to\infty$.
    Thus, $\lim_{V\to\infty}\expv{\trtwo}/V\neq0$ and as such, the moments \labelcref{eq:momentsFull} with $|n|>2$ contain fluctuation contributions, which are not present in \cref{eq:momentsTrOne}.

\section{Chiral Gross-Neveu model}
\label{sec:chiGN}

    As stated before, our representation of the constraint is applicable to any fermionic theory that only features bilinear fermionic terms in its action and thus one could apply it right away to \gls{qcd}.
    However, we first test it in a simpler and computationally less expensive setup, and check whether the approximative form of the constraint still provides a reasonable constraint on the considered fermionic bilinears.
    Arguably, the simplest interacting fermionic theories that feature a spontaneously broken chiral symmetry are low-dimensional four-fermion models, which have a long history serving as toy models in the investigation of various phenomena, see e.g.,
    \Rcite{Schnetz:2005ih,
    Basar:2009fg,
    Gross:1974jv,
    Wolff:1985av,
    Hands:2020itv,
    Lenz:2019qwu,
    Pannullo:2023one,
    Koenigstein:2023yzv}.
    
    The four-fermion model that we consider here as our test case is the \glsfirst{chign} model with $\Nf$ identical flavors of fermions in two space-time dimensions.
    Eventually, we will be interested in the $\Nf\to\infty$ limit, where the mean field approximation becomes exact (see below) and the continuous symmetry is spontaneously broken despite the Mermin-Wagner theorem \cite{Witten:1978qu}.
    The action reads
    \begin{align}
    	S_{\chi\rm GN}={}&-\int \dr[x][2] \Bigg\{ \sum_{f=1}^{\Nf} \bar \psi_f \slashed \partial  \psi_f \\
    	&+ \frac{g^2}{2 \Nf} \left[\left( \sum_{f=1}^{\Nf} \bar \psi_f   \psi_f \right)^2 + \left(\sum_{f=1}^{\Nf} \bar \psi_f \iu \gamma_5  \psi_f \right)^2\right] \vast\}\,, \nonumber
    \end{align}
    where $g^2$ is the coupling constant.
    From here on, we will omit the explicit summation over flavor indices.
    The action is invariant under a chiral symmetry generated by $\gamma_5$, which is broken spontaneously.
    The basis of the order parameter space is given by the bilinears $\m_a$ with $a=0,1$ and the matrices in Dirac space (compare to the definition \labelcref{eq:orderPars}),
    \begin{equation}
    \bilinearMatrix_0=\mathds{1}, \qquad \bilinearMatrix_1=\iu\gamma_5\,.
    \label{eq:generatorscGN}
    \end{equation}
    Note that also on the quantum level, this symmetry is broken only spontaneously, as there is no anomaly that breaks the chiral symmetry in the absence of gauge fields~\cite{Ciccone:2023pdk}.  
    
    The two four-fermion terms in the action are rewritten using the Hubbard-Stratonovich transformation, introducing two bosonic fields $\HS_0$ and $\HS_1$,
    \begin{align}
         \ZGN={}& \int \D\bar\psi\D\psi\, e^{-S_{\chi\rm GN}} \nonumber\\
         ={}& \int \D\HS_a \,\eu^{-N_fS_\HS} \int \D\bar \psi \D \psi \, \exp\left[\bar\psi_f Q\psi_f\right] \nonumber\\
         ={}& \int \D\HS_a \,\eu^{-\Nf S_{\rm eff}[\HS]}\,,
    \label{eq:unconstrZcGN}
    \end{align}
    with the bosonic action
    \begin{equation}
    	S_\HS=\frac{1}{2g^2}\int \dd^2 x  \,\HS_a^2(x)\,,
    \end{equation}
    and the Dirac operator
    \begin{equation}
        Q = \slashed{\partial} + \HS_a \bilinearMatrix_a\,,
    \end{equation}
    entering the effective action
    \begin{equation}
        S_{\rm eff}[\HS]=S_\HS-\log\det Q\,.
    \label{eq:seffsigma}
    \end{equation}
    In \cref{eq:unconstrZcGN} we also performed the fermionic path integral for all $\Nf$ fermion copies.
    In the limit of $\Nf\to\infty$, the path integral collapses to a sum of bosonic field configurations $\HS_a(x)$ that minimize the effective action $S_{\rm eff}$.
    
    Considering \cref{eq:unconstrZcGN} before integrating out fermions, we can perform an infinitesimal variation of the $\HS_a(x)$ field. The invariance of $\ZGN$ under such a variable change gives rise to a Ward identity,
    \begin{equation}
        \frac{\langle\bar\psi_f(x)\bilinearMatrix_a\psi_f(x)\rangle}{\Nf} = 
        \langle\Tr \left(Q^{-1}\bilinearMatrix_a\delta_x\right) \rangle=
        \frac{\langle \HS_a(x) \rangle}{g^2}\, \,,
        \label{eq:WardidcGN}
    \end{equation}
    where $\delta_x$ is a Dirac $\delta$ in coordinate space concentrated at the point $x$.

\subsection{Constrained path integral}

    For the constrained path integral, the Hubbard-Stratonovich transformations can be performed in the same manner, so that we end up with an expression of the general form \labelcref{eq:defZphi},
    \begin{align}
    	\ZGNphi={}& 
    	\int \D\bar\psi\D\psi\, \eu^{-S_{\chi\rm GN}}\,
    	\delta^{(2)}\!\left( \Nf\phi_a - \m_a \right)\nonumber\\
    	={}&
    	\int \D\HS_a \,\eu^{-N_fS_\HS} \label{eq:constrZcGN}\\
    	&\times\int \D\bar \psi \D \psi \, \exp\left[\bar\psi_f Q\psi_f\right]  \delta^{(2)}\!\left( \Nf\phi_a - \m_a \right)\,, \nonumber
    \end{align}
    Notice that we included a normalization factor $\Nf$ in the argument of the Dirac $\delta$ in \cref{eq:constrZcGN} so that $\phi_a$ is ``intensive'' in the flavor number.
    We denote the expectation values of this constrained path integral by $\langle \cdot \rangle_\phi$.
    
    This path integral is now completely analogous to \cref{eq:defZphi}, where the bosonic fields $U$ correspond to the Hubbard-Stratonovich fields $\rho_a$.
    Therefore, we can immediately write down the equivalent of \cref{eq:etaIntegralII},
    \begin{align}
        \ZGNphi={}&\int \D\HS_a \,\eu^{-\Nf S_\HS}\frac{\det^{\Nf} Q}{\sqrt{\det \trtwo}} \label{eq:partfuncnoetacGN}\\ 
        &\times \exp\left[ -\frac{V \Nf}{2}(\phi_a-\trone_a) \trtwo^{-1}_{ab} (\phi_b-\trone_b) \right]\,, \nonumber
    \end{align}
    with the $\trone_a$ and $\trtwo_{ab}$ functionals of the $\HS_a$ fields, this time with the matrices \cref{eq:generatorscGN}.
    The only difference is the additional factor $\Nf$ that appeared in the exponent of \cref{eq:partfuncnoetacGN} due to our slightly different normalization. 
    Due to the large $\Nf$ limit, we again merely need to minimize the effective action, this time of the form
    \begin{equation}
        S_{\rm eff,\phi}[\HS]=S_\HS -\log\det Q + \frac{V}{2}(\phi_a-\trone_a) \trtwo^{-1}_{ab} (\phi_b-\trone_b)\,, \label{eq:chiGNeffectiveactionConstrained}
    \end{equation}
    in place of \cref{eq:seffsigma}.
    The $\sqrt{\det \trtwo}$ factor becomes irrelevant in this limit.
    
    Finally, we point out that the Ward identity still holds in the constrained path integral as
    \begin{align}
         \frac{\langle \HS_c(x) \rangle_\phi}{g^2}  ={}& \frac{\langle\bar\psi_f(x)\bilinearMatrix_c\psi_f(x)\rangle_\phi}{\Nf} \nonumber\\
         ={}& \phi_c +  \Big\langle (\phi_d - \trone_d) \trtwo^{-1}_{da} \trthree_{abc} \trtwo^{-1}_{be} (\phi_e - \trone_e)\Big\rangle_\phi\nonumber\\
         &- \frac{1}{V \Nf}\Big\langle \trtwo^{-1}_{ab} \trthree_{abc} \Big\rangle_\phi, 
     \label{eq:WardidConstraintcGN}
    \end{align}
    where the last equality represents the modified expression for the constrained order parameter as derived in \cref{eq:apprConstraintCondensate}.
    The last term in the expectation value is irrelevant in the large-$\Nf$ limit.
    
    We note that an alternative constraint scheme would be a constraint on the bosonic fields of the form $\delta^{(2)}\!\left( \phi_a - \frac{1}{V} \int \dd^2 x  \,\HS_a(x) \right)$, which could be fulfilled exactly in constrast to the approximated fermionic constraint scheme.
    While the Ward identity \labelcref{eq:WardidcGN} suggests that this constraint would also constrain the fermionic bilinears to the desired values, one finds that a constraint on the bosonic fields modifies the Ward identity with extra terms, which are suppressed in the infinite volume limit.
    An investigation comparing a bosonic constraint and the approximated fermionic constraint in the Gross-Neveu model was presented in \Rcite{Endrodi:2023est}, which found the fermionic constraint to constrain the fermionic bilinears more accurately than the bosonic constraint in finite volumes. 

\subsection{Numerical setup}

    We briefly comment on the numerical setup of our calculation of the constrained \gls{chign} model.
    The model is discretized on a lattice with spacing $a$ and volume $V=L^2=(aN_s)^2$ with
    the fermionic part of the action being formulated using naive fermions.
    The Yukawa interactions in a discretization with fermion doublers are known to lead to interactions between the doublers.
    To suppress these interactions in the continuum limit, we introduce a smearing function $f(p)$ that suppresses the interactions via high momentum components of the bosonic fields.
    We refer to the discussion in \Rcite{Lenz:2020bxk} for details on this discretization in a related model.
    
    The constraint effective action \labelcref{eq:chiGNeffectiveactionConstrained} is minimized using a local minimization algorithm in the field variables.
    To ensure that the found minimum is the global one, we start the minimization from several starting configurations.
    The employed code and setup have been used in \Rcite{Pannullo:2021edr,Winstel:2022jkk,Pannullo:2024sov}, which also contain further information on the details.

\section{Numerical results in the constrained chiral Gross-Neveu model}
\label{sec:results}

    In this section, we present the results of our numerical simulations of the \gls{chign} model with a fermionic constraint.
    In this test, we consider various values of the coupling, i.e., lattice spacings $a$, various lattice extents $\Ns$ and constraining parameter $\phi_0$.
    We set $\phi_1=0$ for the following results. This is an arbitrary choice without loss of generality, since the constraint potential is rotationally symmetric in the $(\phi_0,\phi_1)$ plane due to the chiral symmetry of the unconstrained model.
    The range of parameters is summarized in \cref{tab:parameterTable}.
    We show results at a single lattice spacing for different physical volumes in \cref{sec:resCondensate,sec:resPotential,sec:chiGN_configs,sec:resSaddlePoint} and explore the lattice spacing dependence of our results in \cref{sec:resSpacing}.
    
    To compare results of different lattice spacing values, one typically sets the scale with the expectation value of the $\rho_0$ field extrapolated to the chiral limit, i.e., $\lim_{m \to 0} \lim_{V\to\infty}\langle \rho_0 \rangle_m\equiv \vacScale$, as it has a finite continuum limit. 
    While the determination of this quantity can be costly in a Monte Carlo simulation, it is relatively inexpensive to calculate this in our large-$\Nf$ calculations for all lattice spacings and extents that we considered.
    Via the Ward identity \cref{eq:WardidcGN}, one can then identify $\phimin = \vacScale / g^2$, which is the symmetry breaking value of the chiral condensate.
    The goal of our investigation is to demonstrate that the value of this observable coincides with the position of the edge of the flat disk in $\Omega(\phi)$ in the thermodynamic limit, i.e., our alternative approach via the constrained path integral yields the same result.
    
    \begin{table}[]
        \centering
        \begin{tabular}{c|cccc}
             $g^2$& $a\vacScale$ &$\Ns$ & $L\vacScale$ & $\phi_0 / \phimin$ \\\hline
             $0.492029$& $1/2$ &20,28,40,60& 10,14,20,30 & [0.0,...,1.2] \\
             $0.369784$& 0.3125 &32  &10 & [0.0,...,1.5] \\
             $0.331692$& $1/4$  &40  &10 & 0.0,0.5, [1.0,...,1.5] \\   
             $0.280538$& $1/6$  &60  &10 & 0.0,0.5, [1.0,...,1.5] \\            
             $0.223494$& $1/12$ &120 &10 & 0.0,0.5, [1.0,...,1.5] \\    
             $0.186354$& $1/24$ &240 &10 &  [1.0,...,1.5] \\  
        \end{tabular}
        \caption{Summary of the employed simulation parameters.}
        \label{tab:parameterTable}
    \end{table}

\subsection{The constrained condensate}
\label{sec:resCondensate}

    \begin{figure*}
        \centering
        \includegraphics{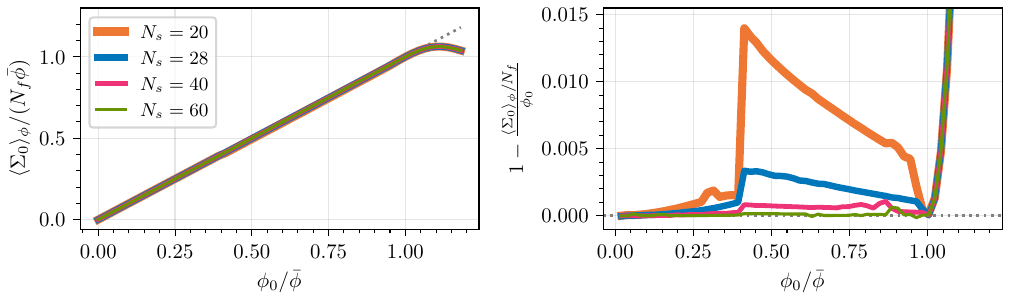}
        \caption{The expectation value of the fermionic condensate $\langle\m_0\rangle_{\phi}$ as a function of the constraint $\phi_0$ in the left plot and the relative deviation from $\phi_0$ in the right plot for the lattice sizes $\Ns=20,28,40,60$ and $\phi_1=0$. }
        \label{fig:chiGN_cc_vs_constraint}
    \end{figure*}
    
    We start by showing the constrained expectation value of the fermionic condensate $\langle\m_0\rangle_\phi$ calculated according to \cref{eq:WardidConstraintcGN} as a function of the constraint $\phi_0$ in the left panel of \cref{fig:chiGN_cc_vs_constraint} and the relative deviation from $\phi_0$ in the right panel.
    Note that we rescale $\phi_0$ as $\phi_0 / \phimin$, to facilitate a comparison between different lattice spacings as is done in \cref{sec:resSpacing}.
    For comparability, we apply this scaling also in this section, where only a single lattice spacing is considered.
    The expectation value of the condensate is fixed to the constraint value $\phi_0$ remarkably well for small values of $\phi_0$.
    This shows that the correction to $\phi_0$ given in \cref{eq:WardidConstraintcGN} is indeed suppressed for small values of $\phi_0$.
    At an intermediate value of $\phi_0$, the expectation value of the condensate deviates from the constraint value in a jump, which decreases with increasing volume.
    For large values of $\phi_0$, the expectation value of the condensate deviates significantly from the constraint value. These two types of behavior have very different origins, being related to the characteristic change of minimizing field configurations and discretization artifacts, respectively. We get back to these features below.

\subsection{The constraint effective potential}
\label{sec:resPotential}

    \begin{figure*}
        \centering
        \includegraphics[]{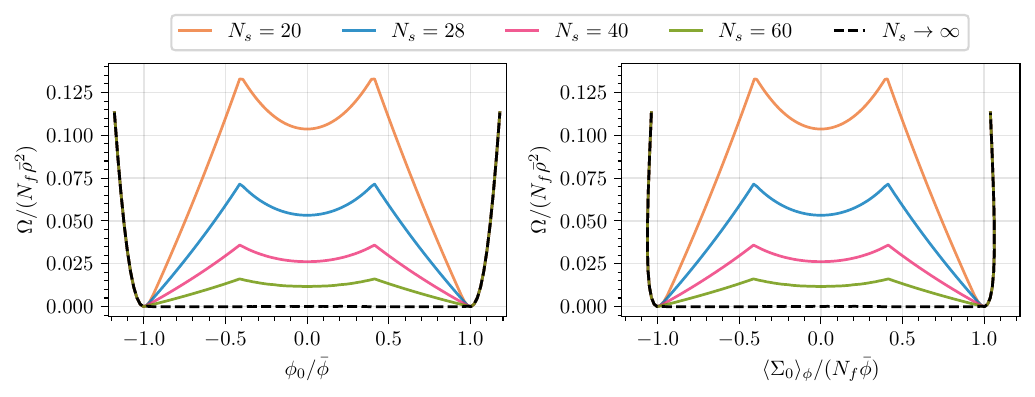}
        \caption{The constraint effective potential $\Omega$ as a function of the constraint $\phi_0$ in the left plot and plotted against the expectation value of the fermionic condensate $\langle\m_0\rangle_{\phi}$ in the right plot for the lattice sizes $\Ns=20,28,40,60$ and $\phi_1=0$. The dashed black line is obtained as the infinite volume extrapolation by fitting $f(\Ns)=\alpha+\beta/\Ns^2+\gamma/\Ns^4$ for every $\phi_0$ individually.}
        \label{fig:chiGN_Omega}
    \end{figure*}
    
    \begin{figure}
        \centering
        \includegraphics[]{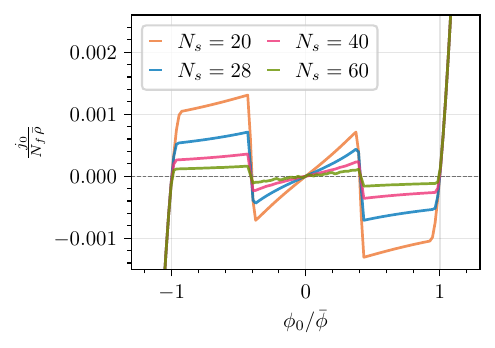}
        \caption{
        The derivative of the constraint potential $\constraintDerivative_0 = \partial\Omega/\partial\phi_0$ as a function of $\phi_0$ for various volumes at fixed lattice spacing.
        }
        \label{fig:chiGN_Omega_deriv}
    \end{figure}
    
    \cref{fig:chiGN_Omega} shows the constraint effective potential $\Omega$ as a function of the constraint $\phi_0$ in the left panel and against the expectation value of the fermionic condensate $\langle\m_0\rangle_{\phi}$ in the right panel for the lattice sizes $\Ns=20,28,40,60$ and $\phi_1=0$.
    The region for $|\phi_0|<\phimin$ flattens for increasing volumes as shown by the infinite volume extrapolation given by the black dashed line.
    For finite volumes, this region features a cusp and non-monotic behavior at $|\phi_0|/\phimin \approx 0.4$.
    This is also the point where the expectation value of the condensate deviates from the constraint value in a jump, see the right panel of \cref{fig:chiGN_cc_vs_constraint}.
    Interestingly, the shape of the inner part is significantly different from the constraint potential of the related \gls{gn} model, which features only a discrete chiral symmetry, that was presented in \Rcite{Endrodi:2023est}.
    On the other hand, its shape is reminiscent of the constraint potential of a bosonic model featuring an $\mathrm{O}(2)$ symmetry as obtained from constrained Monte Carlo simulations discussed in \Rcite{Endrodi:2021kur}.
    This underlines that the constraint potential is mainly shaped by the realized symmetries of the theory rather than its microscopic formulation.
    
    For all volumes, the constraint potential is found to exhibit a minimum at $\phi_0/\phimin \approx1$, or, equivalently, a zero of its derivative, see \cref{fig:chiGN_Omega_deriv}.
    This point marks the edge of the spontaneously broken disk and by taking the infinite volume limit of its position, we can then determine the physically realized expectation value $\langle \m_0 \rangle$ of the unconstrained system.
    Such a procedure of determining the order parameter is applicable in the same way in full Monte Carlo simulations of the \gls{chign} or \gls{qcd}.
    The clear advantage -- as demonstrated here -- is that we do not need to carry out the intricate double limit~\labelcref{eq:standardapproach}, but only a single infinite volume extrapolation. 
    Outside of the valley, $\Omega(\phi)$ increases and continues to do so in the infinite volume limit.
    When plotted against the expectation value of the condensate, one finds a back-bending behavior in this region.
    This is due to the back-bending of the constrained condensate for large values of $\phi_0$ as shown in the left plot of \cref{fig:chiGN_cc_vs_constraint}, which is a lattice discretization artifact as is demonstrated in \cref{sec:resSpacing}.

\subsection{Inhomogeneous condensates in the flattening region}
\label{sec:chiGN_configs}
    
    In the previous sections, we observed that the expectation value of the fermionic condensate is fixed to the constraint value up to a small jump, and the constraint potential shows a cusp at this point and features non-monotonic behavior.
    In this section, we explain these effects by showing the spatial dependence of the fermionic condensates in the flattening region.
    
    As discussed in \cref{sec:constrpot}, the constrained path integral is dominated by inhomogeneous configurations inside the valley. We find that these inhomogeneities resemble spin-waves, just like in the $\mathrm{O}(2)$ model~\cite{Endrodi:2021kur}.
    These configurations always exhibit a marked dependence on one of the space-time coordinates and are constant in the other direction.
    Thus, they lead to the spontaneous breaking of translational and rotational symmetry of the theory, similarly to the inhomogeneous phases found at nonzero chemical potential in this model~\cite{Basar:2009fg}.
    For visualization purposes, below we concentrate on the nontrivial local dependence of the condensate and average in the trivial direction, but simply denote it as $\expv{\m_a(x)}$.
    
\begin{figure*}
        \centering
        \includegraphics[width=1.0\linewidth]{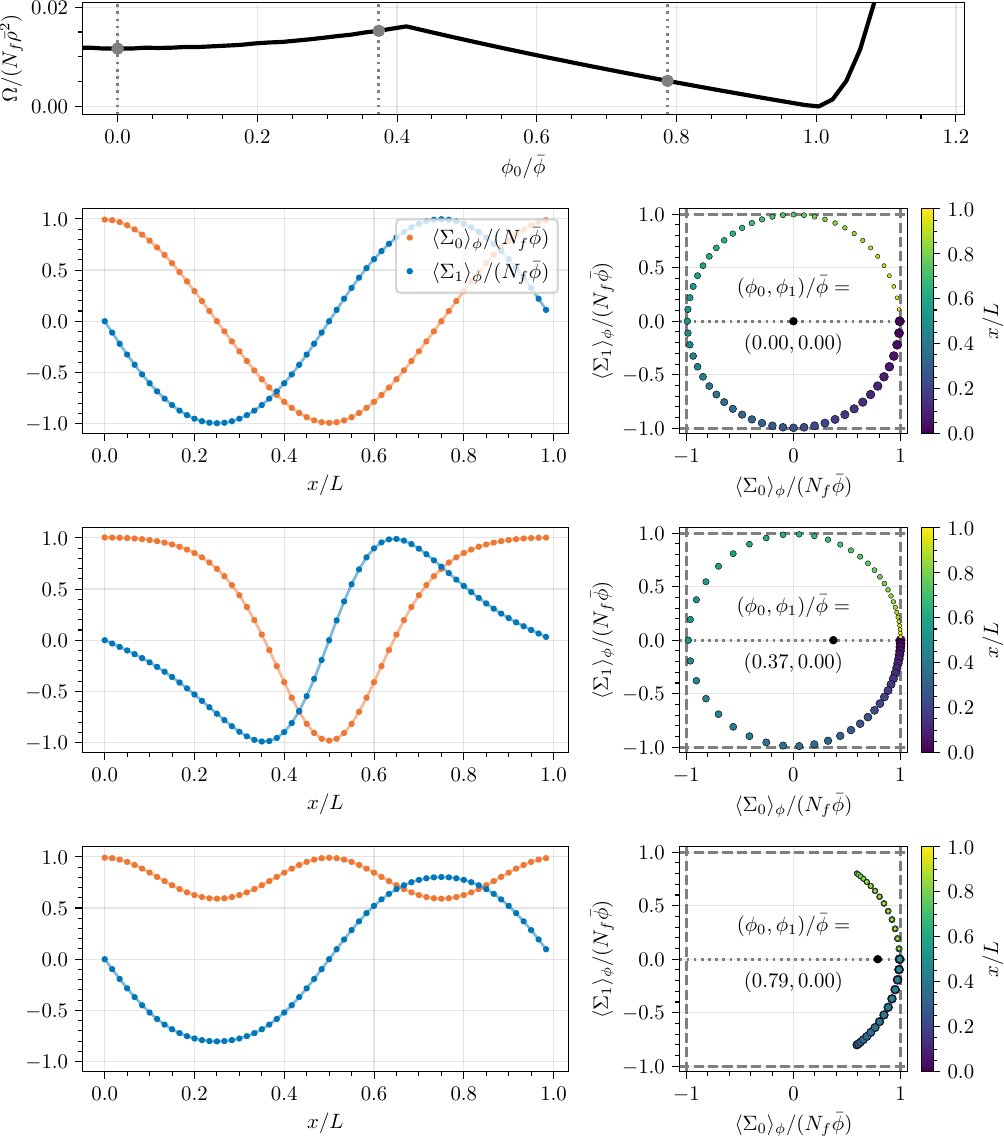}
        \caption{The spatially dependent condensates $\langle\m_0(x)\rangle_{\phi},\langle\m_1(x)\rangle_{\phi}$ for the lattice size $\Ns=60$ and the constraint values $\phi_0/\phimin=0.0,0.37,0.79$, $\phi_1=0$.
    	The top plot shows the corresponding position in the constraint effective potential of the three rows below, where each one corresponds to a different constraint value.
    	In each row, the left plot shows the condensates as a function of space, the right plot shows the condensates in their combined plane, where the color bar and marker size indicates the spatial position.}
        \label{fig:chiGN_field_configs}
    \end{figure*}

    \cref{fig:chiGN_field_configs} shows the spatial dependence of the condensates $\expv{\m_0(x)}_\phi$ and $\expv{\m_1(x)}_\phi$ for the lattice size $\Ns=60$ and the constraint values $\phi_0/\phimin=0,0.37,0.79$.
    The top panel indicates the constraint parameters and the corresponding values of the constraint effective potential $\Omega(\phi)$.
    Each of the lower rows corresponds to a single constraint value $\phi_0$.
    The left panels show the condensates as a function of space, while the right panels the condensates in the $(\m_0,\m_1)$ plane, where the color bar and marker size indicates the spatial position.
    
    Thus, we find that the flattening region of the constraint potential is associated with in\-ho\-mo\-ge\-ne\-ous field configurations.
    For small values of $\phi_0$, one finds that the condensates fully wind around the valley of the potential.
    As $\phi_0$ increases, this winding tends to occur with a non-uniform ``velocity'' in order to fulfill the constrained average value.
    This increases the kinetic energy of the condensate, which is reflected in increasing values of the constraint potential.
    In turn, the cusp in the constraint potential is associated with a change of the winding behavior of the condensates.
    For values of $\phi_0$ larger than the cusp, the condensates exhibit a non-winding behavior, which is energetically favorable compared to the full winding.
    Increasing the constraint value further, one finds that the condensates cover a decreasing angle interval in the valley and approach homogeneous configurations.
    This kind of behavior is completely analogous to what was observed for the bosonic field in the constrained $\mathrm{O}(2)$ model~\cite{Endrodi:2021kur}.
    Note that the occurrence of winding configurations is due to the use of periodic boundary conditions.
    For example, with open boundary conditions one might expect configurations that cover arcs of the valley of the potential akin to the non-winding configurations that we find, but without the turning back.
    
    In the \gls{gn} model with a discrete chiral symmetry, it was found that the flattening region of the constraint potential is also dominated by inhomogeneous field configurations, but they tunnel through the potential barrier between the two minima \cite{Endrodi:2023est}.
    This is due to the discrete symmetry, which does not allow the kind of behavior that was observed here with the continuous symmetry.

\subsection{The saddle-point approximation on the minimizing configurations}
\label{sec:resSaddlePoint}

\begin{figure*}[p]
        \centering
        \includegraphics[width=0.42\linewidth,clip,trim= 3mm 3mm 0mm 7mm]{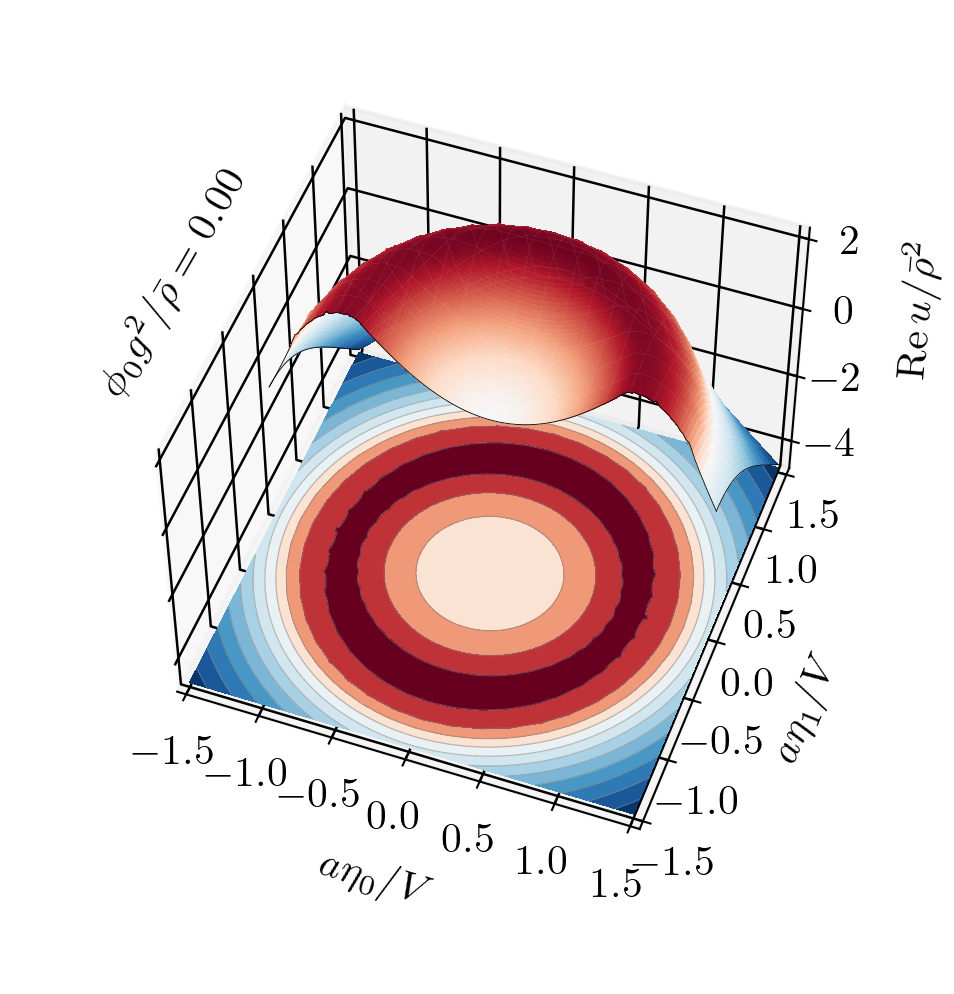}
        \hfill
        \includegraphics[width=0.42\linewidth,clip,trim= 3mm 3mm 0mm 7mm]{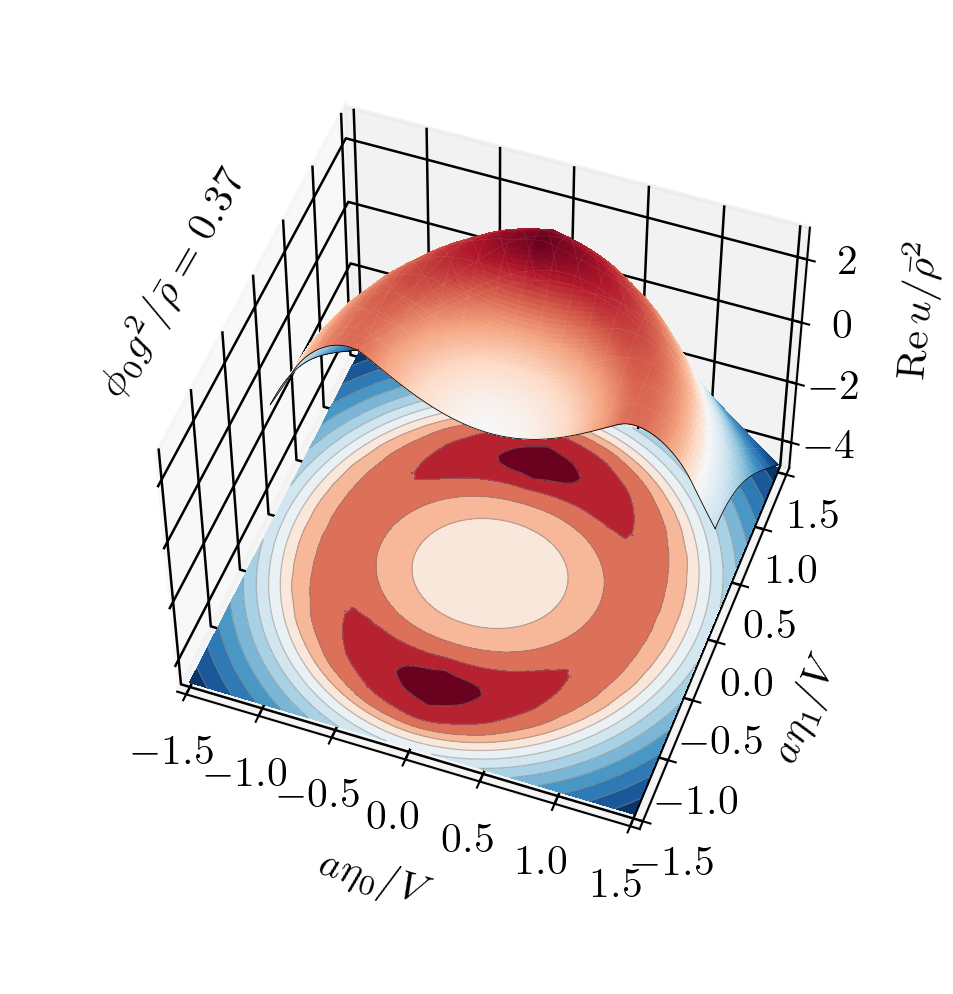}\\[3mm]
        \includegraphics[width=0.42\linewidth,clip,trim= 3mm 3mm 0mm 7mm]{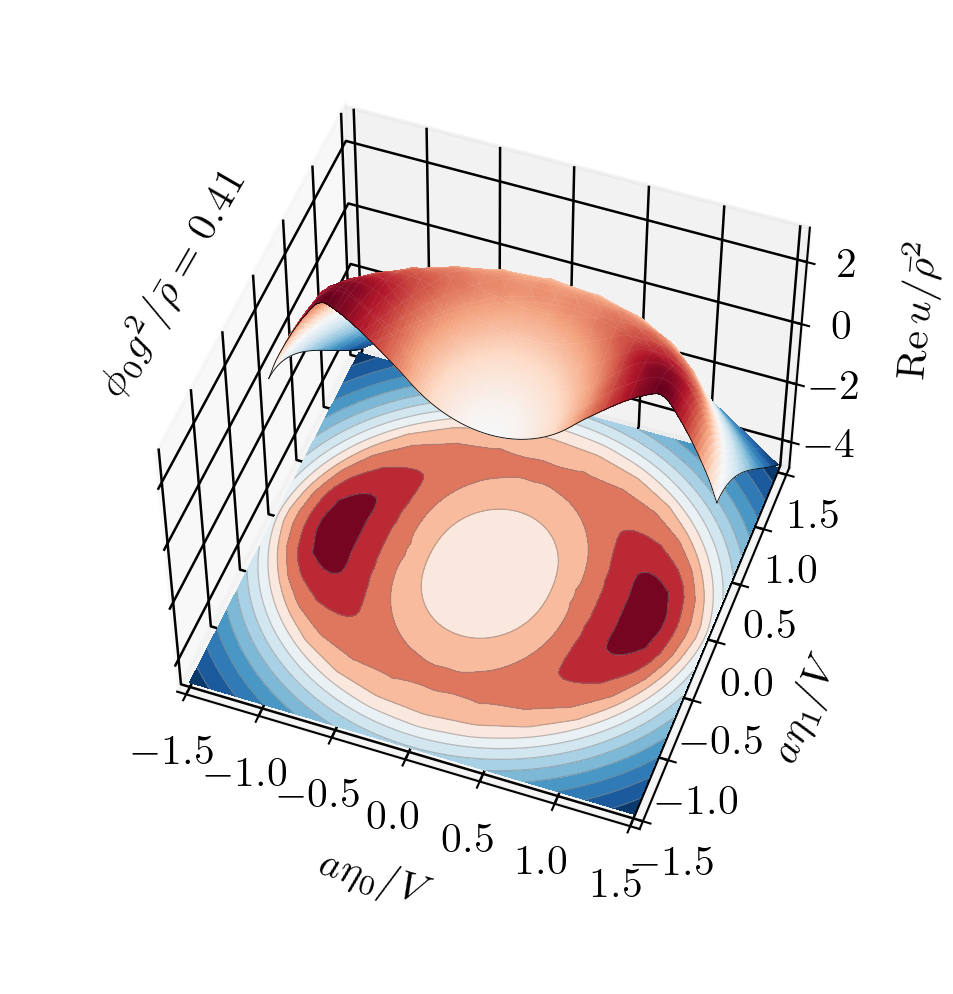}
        \hfill
        \includegraphics[width=0.42\linewidth,clip,trim= 3mm 3mm 0mm 7mm]{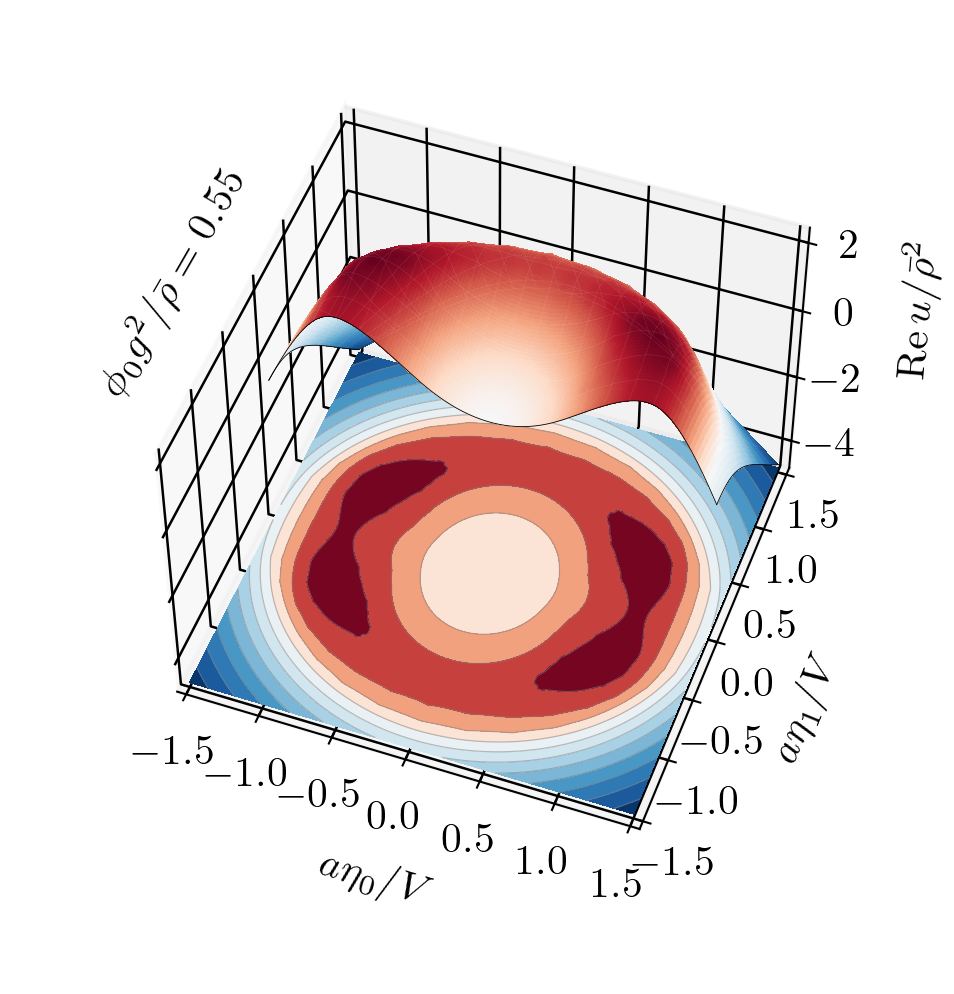}\\[3mm]
        \includegraphics[width=0.42\linewidth,clip,trim= 3mm 3mm 0mm 7mm]{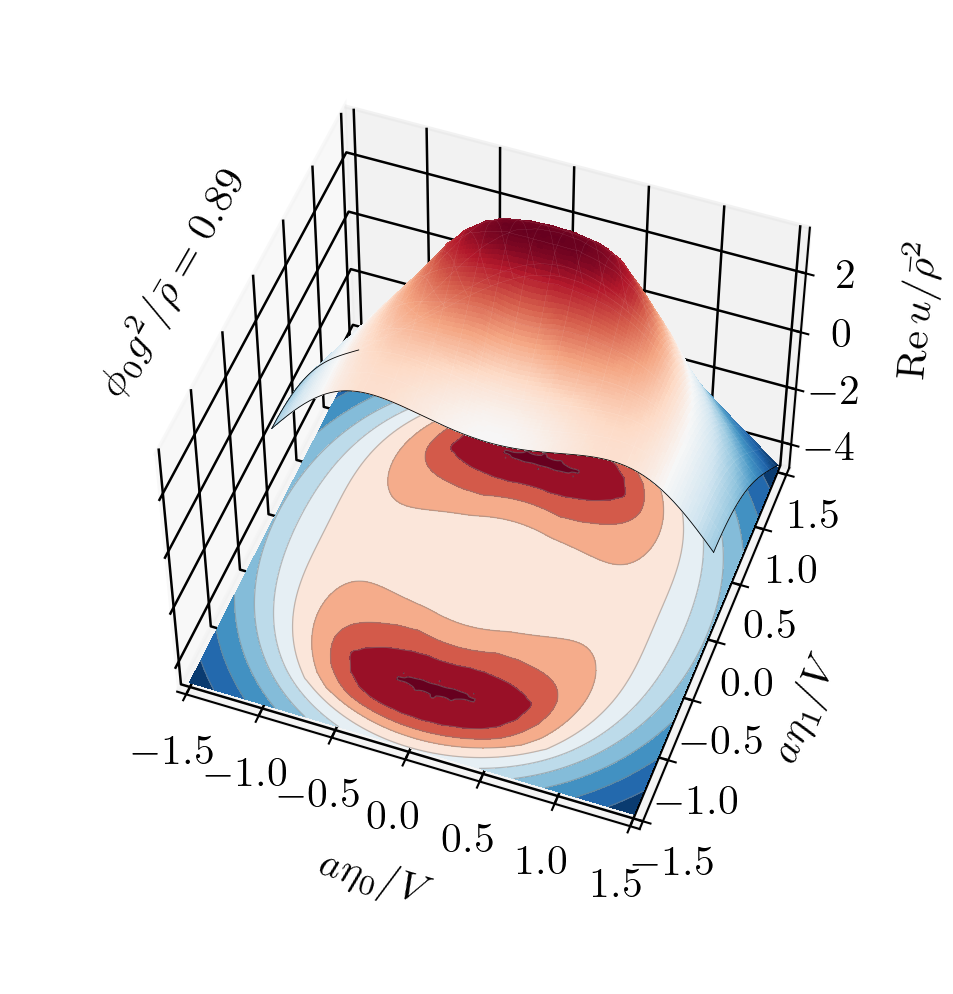}
        \hfill
        \includegraphics[width=0.42\linewidth,clip,trim= 3mm 3mm 0mm 7mm]{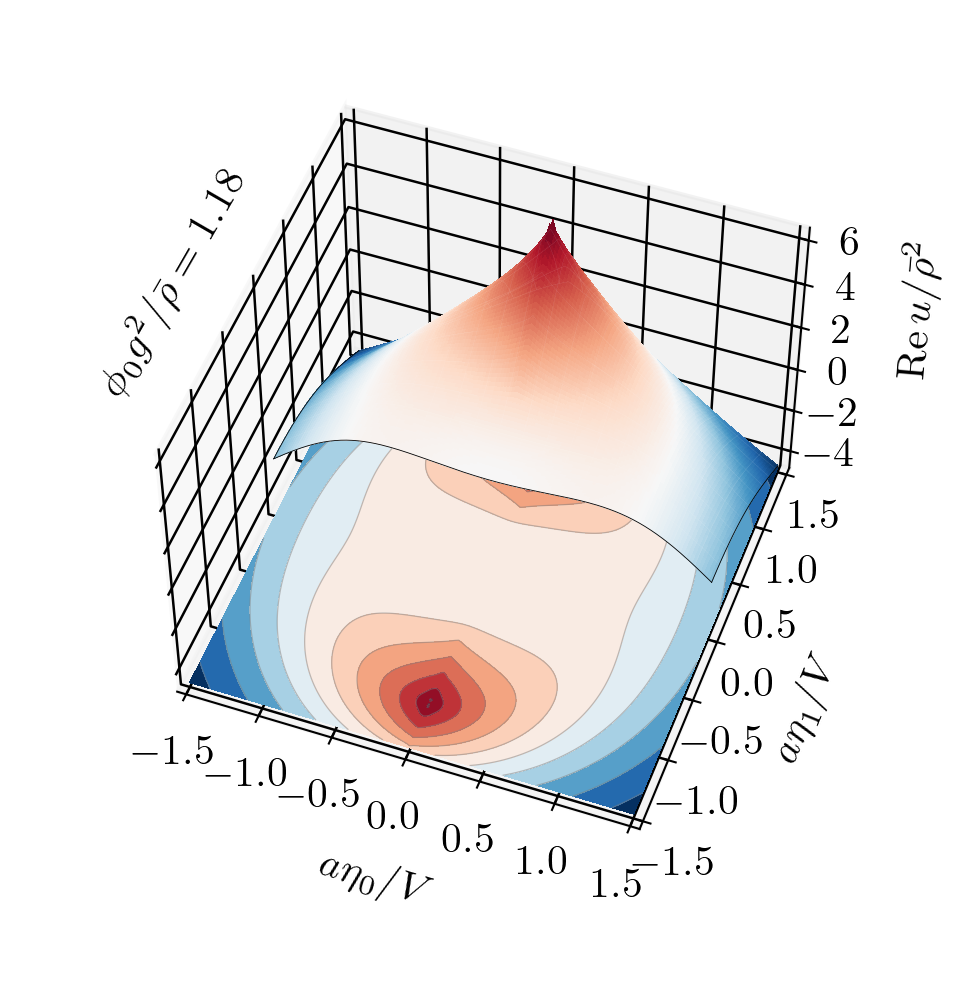}
        \caption{$ \mathrm{Re}\, \uSmall$ in the plane of $\DDR_0,\DDR_1$ evaluated on $\rho$ field configurations that are minimzing field configuration of \cref{eq:chiGNeffectiveactionConstrained} for $\Ns=40$ at the respective values of $\phi_0$ and $\phi_1=0$. For this plot, $\uSmall$ is normalized such that $ \mathrm{Re}\, \uSmall (\DDR_0=0,\DDR_1=0)=0$. 
        The contourplot at the bottom of the 3D axis shows $ \mathrm{Re}\, \uSmall$ with blue being negative values and red positive values.
        The 3D surface shows the same data in the half plane of positive $\eta_1$.}
        \label{fig:etaScanU}
    \end{figure*}

    This section aims to illustrate the impact of the saddle-point approximation of $\uSmall$ in $\DDR$, which is the crucial step in our implementation of the constraint (see \cref{sec:ApproximateSaddlepointApproximation}).
    To do so, we calculate $\uSmall$ on the field configurations that were obtained from the minimization of \cref{eq:chiGNeffectiveactionConstrained} on the $\Ns=40$ lattice, i.e., the configurations that the path integral reduces to in the $\Nf\to\infty$ limit.
    In this way we can confirm that the saddle-point approximation is valid at the very least for these configurations.
    \cref{fig:etaScanU} shows $\Real \uSmall$ evaluated on these configurations for various $\phi_0$ and $\phi_1=0$.
    These figures demonstrate that there is always a well defined local minimum in $\Real \uSmall$ at $\DDR_0=\DDR_1=0$ for the relevant field configurations allowing an expansion about $\DDR_0=\DDR_1=0$ as motivated in \cref{sec:ApproximateSaddlepointApproximation}. 
    Recall that in a finite volume the true saddle-point is located at $\Real \DDRExtreme=0$ and $\Imag \DDRExtreme\neq0$, which would be the correct expansion point in a saddle-point expansion.
    However, for $|\phi|<\phimin$ and in the infinite volume limit one finds $\lim_{V\to\infty}\DDRExtreme=0$ recovering the exactness of our expansion.
    
    We notice that $\uSmall$ negatively diverges for large $|\DDR/V|\gtrsim a^{-1}$.
    There, $\DDR$ becomes the largest scale and dominates the value of $\uSmall$, which is a lattice artifact.
    This is akin to the situation where bare quark masses are comparable to the inverse lattice spacing.
    Thus, by carrying out the saddle point approximation, we suppress these artifact contributions that would be present in the full integral over $\DDR$.
    
    Furthermore, there is a jump in the overall shape of $\uSmall$ when going from $\phi_0/\phimin=0.37$ to $0.41$, which is where the change from winding to non-winding condensate configurations occurs.
    
\subsection{Lattice spacing dependence}	
\label{sec:resSpacing}

    In this section, we show that the results that we presented at a fixed lattice spacing in the previous sections, are qualitatively similar across various lattice spacings and not the result of discretization artifacts.
    
\begin{figure*}
        \centering
        \includegraphics{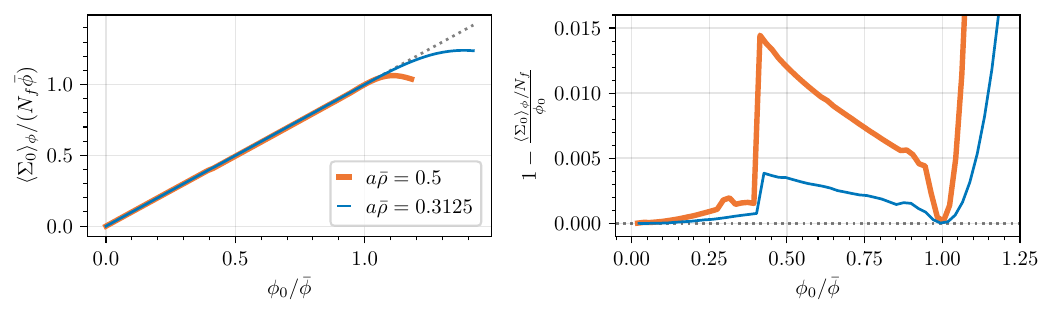}
        \caption{The chiral condensate $\langle \m_0 \rangle_{\phi}$ (left) and its relative deviation from $\phi_0$ (right) as a function of $\phi_0$ for two lattice spacings at fixed physical volume $L\vacScale=10$. }
        \label{fig:chiGN_cc_vs_constraint_latticeSpacings}
    \end{figure*}

    \cref{fig:chiGN_cc_vs_constraint_latticeSpacings} shows the chiral condensate $\langle \m_0 \rangle_{\phi}$ and its relative deviation from $\phi_0$ as a function of $\phi_0$ for two lattice spacings at fixed physical volume $L\vacScale=10$.
    We note that the deviation of $\langle \m_0 \rangle_{\phi}$ from $\phi_0$ decreases in the whole range of $\phi_0$.
    Especially the back-bending in the region of $\phi_0 > \phimin$ moves to larger $\phi_0$ for decreasing lattice spacing.
    
\begin{figure}
        \centering
        \includegraphics{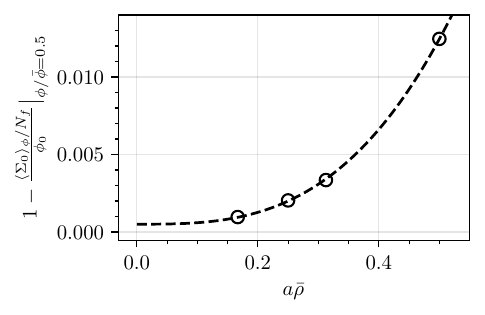}
        \caption{The relative deviation of the chiral condensate $\langle \m_0 \rangle_{\phi}$  from $\phi_0$ at $\phi_0/\phimin=0.5$ as a function of the lattice spacing at fixed physical volume $L\vacScale=10$ and an extrapolation to the continuum obtained via a fit of the function $f(a\vacScale) = \alpha (a\vacScale)^3 + \beta $. }
        \label{fig:chiGN_cc_vs_latticespacing_midpoint}
    \end{figure}

    \cref{fig:chiGN_cc_vs_latticespacing_midpoint} shows the value of the relative deviation of the chiral condensate $\langle \m_0 \rangle_{\phi}$  from $\phi_0$ at $\phi_0 /\phimin=0.5$ as a function of the lattice spacing and an extrapolation to the continuum obtained via a fit of the function $f(a\vacScale) = \alpha (a\vacScale)^3 + \beta $ with an obtained value of $\beta=0.000503$.
    This shows that also in the flattening region, lattice artifacts play a sizeable role at a finite volume in the deviation and that the deviation assumes a finite albeit small value in the continuum limit.
    Due to our findings in \cref{{sec:resCondensate}} we expect the residual deviation to vanish in the infinite volume limit. 
    
\begin{figure}
        \centering
        \includegraphics{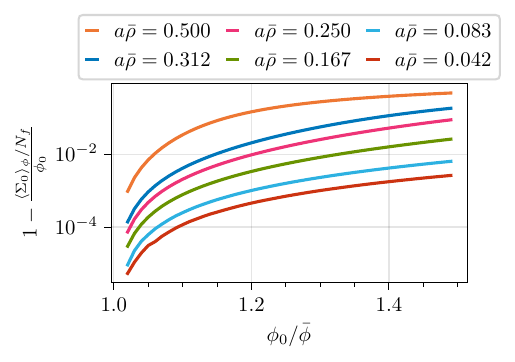}
        \caption{The relative deviation of the chiral condensate $\langle \m_0 \rangle_{\phi}$  from $\phi_0$ at $\phi_0/\phimin=0.5$ as a function of $\phi_0$ for various lattice spacings at a fixed physical volume $ L\vacScale=10$.}
        \label{fig:chiGN_cc_vs_vs_phi_outer_latticespacings}
    \end{figure}

    In the region outside the disk, we observed that the deviation persists in the infinite volume (see \cref{sec:resCondensate}).
    \cref{fig:chiGN_cc_vs_vs_phi_outer_latticespacings} shows that a finite deviation for any $\phi_0 > \phimin$ remains at various lattice spacing.
    They are, however, small and close to the precision of our minimization procedure in these calculations.
    
\begin{figure*}
        \centering
        \includegraphics{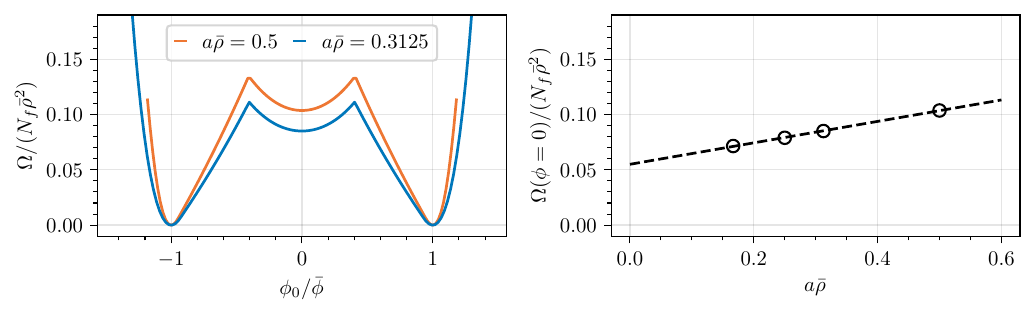}
        \caption{The constraint potential as a function of $\phi_0$ for two lattice spacings (left) and the central value of the constraint potential $\Omega(\phi_0=0)$ as a function of the lattice spacing (right) both at a fixed physical volume $L\vacScale=10$.}
        \label{fig:chiGN_Omega_vs_constraint_latticeSpacings_combined}
    \end{figure*}

    Lastly, we show the constraint potential for two different lattice spacings in the left plot of \cref{fig:chiGN_Omega_vs_constraint_latticeSpacings_combined}. 
    The potential reduces in magnitude for all $\phi_0$, but converges to a finite value in the continuum limit as illustrated in the right plot at the example of $\Omega(\phi_0=0)$.
    
    In summary, we find that discretization artifacts are mild and do not substantially change the results.
    
\section{Summary, Conclusion and Outlook}
\label{sec:conclusion}

    In this work, we have presented a novel method to implement a constraint on fermionic condensates ${\m_a=\bar \psi \bilinearMatrix_a \psi}$ to obtain a fermionic constrained path integral.
    This gives access to the constraint effective potential and, thus, in the infinite volume limit to the quantum effective potential of the theory.
    Due to the Grassmann nature of the fermionic fields, the realization of the constraint is non-trivial.
    Our formulation of the constraint is implemented by the saddle-point approximation and large-volume expansion of the exact constraint. We have shown that the approximated constraint method constitutes a constrained path integral that matches the correct moments of the exact constrained path integral up to finite volume artifacts.
    
    We have demonstrated the method in the \glsfirst{chign} model -- a four-fermion model with a spontaneously broken continuous chiral symmetry -- in the large-$N_f$ limit.
    Our results show that the constrained fermionic condensates are fixed to the constraint value up to finite volume corrections. 
    The constraint effective potential exhibits a non-trivial shape with a region that flattens in the infinite volume limit approaching the quantum effective potential of the unconstrained theory.
    This flattening region is associated with inhomogeneous condensates that break translational and rotational symmetry of the theory.
    We have shown that lattice discretization errors in the constraint effective potential and the constrained condensates are mild.
    
    The method presented in this work is general and not limited to low-dimensional models or the large-$N_f$ limit.
    It can be applied to any fermionic theory with a bilinear action $\bar \psi Q \psi$ by a modification of the Monte Carlo simulations -- in particular \gls{qcd}.
    In the future, such constrained simulations should be implemented in \gls{qcd} to study its quantum effective potential, enabling us to study the chiral limit of the theory in an alternative way (compared to the standard approach of the double limit) and to investigate the nature of the chiral anomaly.
	
\acknowledgments 

    L.P.~acknowledges interesting and helpful discussions with A.~Koenigstein, M.~Winstel and A.~Wipf.
    We would like to thank B.~Brandt for the helpful discussions.
    This work received funding from the DFG (Collaborative Research Center CRC-TR 211 ``Strong-interaction matter under
    extreme conditions'' - project number 315477589 - TRR 211) and the Hungarian National Research, Development and Innovation Office (Research Grant Hungary). T.G.K.\ was supported by NKFIH grant K-147396 and excellence grant TKP2021-NKTA-64. 
	
\appendix

	\section{Interpretation of the Dirac \boldmath$\delta$ \label{app:foolsGold}}
	In this appendix, we will discuss the formal meaning of the constraint $\delta(\m_a-\phi_a)$ for the case of a finite number of fermionic degrees of freedom, i.e.\ $\psi_n$ and $\bar\psi_n$ with $1\le n\le N$, corresponding to a lattice discretized theory in a finite volume. 
	For simplicity, first we consider only one component $a$ and generalize to an arbitrary number of constraints later.	
	
	First of all, we point out that $\m_a-\phi_a$ is a so-called supernumber, discussed usually within the field of superanalysis. While this field developed enormously thanks to supersymmetry, we only need to use the basics here. The formal meaning of a Dirac $\delta$	can be found e.g.\ in the first chapter of \Rcite{DeWitt:2012mdz}. It is essentially the generalization of the function definition for Grassmann numbers, which uses their naturally truncated Taylor expansion. In our case, this natural truncation occurs due to $\m_a^{N+1}=0$. Formally applying the Taylor expansion to the Dirac $\delta$ yields
	\begin{equation}
	    \begin{split}
	    \label{eq:deltaDeWitt}
	    \delta(\phi_a-\m_a) &= \sum_{n=0}^\infty \frac{(-1)^n}{n!}\m_a^n\delta^{(n)}(\phi_a)\\
	    &= \sum_{n=0}^{N}\frac{(-1)^n}{n!}\m_a^n\delta^{(n)}(\phi_a)\,,
	    \end{split}
	\end{equation}
	where we used the notation $\delta^{(n)}$ for the $n$-th derivative of the $\delta$ distribution. It is easy to see that applying this formula for the constrained path integral~\labelcref{eq:defZphi} leads to a (generally large) sum of distributions,
	\begin{equation}
	    \label{eq:foolsGold}
	    \Zphi = \Z \sum_{n=0}^{N}\frac{(-1)^n}{n!}\expv{\m_a^n}\delta^{(n)}(\phi_a)\,,
	\end{equation}
	using which a constraint potential, $\Omega(\phi)\sim\log\Zphi$ cannot be defined in practice.
	
	An alternative, equivalent representation of the $\delta$ can be constructed by considering its Fourier series. For ordinary variables
	\begin{equation}
	    \delta(x) = \lim_{\varepsilon\to0}\int_{-\infty}^\infty \frac{\dd\DDR}{2\uppi}\,\eu^{i\DDR x-\DDR^2\varepsilon/(2V)}\,,
	\end{equation}
	where $\Real{\varepsilon}>0$ to ensure convergence. Since the exponential function is well defined for $\m_a$ through its Taylor series, this can be generalized to $x=\phi_a-\m_a$ straightforwardly, resulting in,
	\begin{equation}
	    \label{eq:deltawFT}
	    \delta(\phi_a-\m_a) = \lim_{\varepsilon\to0}\int_{-\infty}^\infty \frac{\dd\DDR_a}{2\uppi}\,\eu^{i\DDR_a (\phi_a-\m_a)-\DDR_a^2\varepsilon/(2V)}\,,
	\end{equation}
	where we introduced the index $a$ on $\DDR$ for later convenience. 
	
	Expanding $\eu^{-i\DDR_a\m_a}$ in its Taylor series and remembering that
	\begin{equation}
	     \lim_{\varepsilon\to0}\int_{-\infty}^\infty\frac{\dd\DDR_a}{2\uppi} (-\iu\DDR_a)^n \eu^{i\DDR_a\phi_a-\DDR_a^2\varepsilon/(2V)} = (-1)^n\delta^{(n)}(\phi_a)\,,
	\end{equation}
	immediately recovers \cref{eq:deltaDeWitt}. But for us the more important aspect of \cref{eq:deltawFT} is that due to the bilinearity of $\m_a$ in the fermion fields, the fermionic path integral can be carried out if one exchanges it with the $\DDR_a$ integral. Recalling the fermionic part of the action from \cref{eq:fermionAction}, we find that
	\begin{equation}
	    \label{eq:ZphiWeta}
	    \begin{split}
	    \Zphi={}&\int\D U\D\bar\psi \D\psi \, \eu^{-\Sb[U]+S_{\rm F}[\bar\psi,\psi,U]}\,\delta(\phi_a-\m_a)\\ ={}&\lim_{\varepsilon\to0}\int \D U\eu^{-\Sb[U]}\\
	    &\times\int_{-\infty}^\infty\frac{\dd\DDR_a}{2\uppi} \det\left(Q-i\frac{\DDR_a\sigma_a}{V}\right)\eu^{i\DDR_a\phi_a-\DDR_a^2\varepsilon/(2V)}\,,
	    \end{split}
	\end{equation}
	where no summation over $a$ is implied. At this point, carrying out the $\DDR_a$ integral analytically is not possible anymore. However, to recover \labelcref{eq:foolsGold} one can expand the determinant in a 
	Taylor series in $\DDR_a$ around zero, remembering that it is a finite polynomial in $\DDR_a$ owing to the finiteness of the fermionic degrees of freedom. Therefore
	\begin{equation}
	    \det\left(Q-i\frac{\DDR_a\sigma_a}{V}\right) = \sum_{n=0}^N\frac{\DDR_a^n}{n!}\frac{\partial^n}{\partial \DDR_a^n}\det\left(Q-i\frac{\DDR_a\sigma_a}{V}\right)\Big|_{\DDR_a=0}\,.
	\end{equation}
	Each derivative results in a constant factor of $-\iu/V$, and setting $\DDR_a=0$ after the differentiations practically decouples the $\DDR_a$ integral from the rest. The remaining part corresponds to the $n$-th derivative of the unconstrained fermionic path integral with respect to a source, which generates exactly $\expv{\m_a^n}$. With that we find
	\begin{equation}
	    \label{eq:TaylorPIdone}
	    \Zphi = \int_{-\infty}^\infty \frac{\dd\DDR_a}{2\uppi}\sum_{n=0}^N \frac{(-1)^n\expv{\m_a^n}}{n!}\left(\frac{\iu \DDR_a}{V}\right)^n \eu^{\iu\DDR_a\phi_a-\DDR_a^2\varepsilon/(2V)}\,.
	\end{equation}
	In this form the $\DDR_a$ integral can again be carried out and \cref{eq:foolsGold} is recovered. 
	
	The generalization of \cref{eq:ZphiWeta} to include constraints in several directions is straightforward, practically amounting to including new $\DDR_a$ integrals for each $\delta$ and summing over repeating $a$-indices. The results obtained show that even for finite many fermionic degrees of freedom it is meaningful to define the object
	\begin{equation}
	    \Zeta=\int \D U \eu^{-\Sb[U]}\det\left(Q-i\frac{\DDR_a\sigma_a}{V}\right)\,,
	\end{equation}
	where a summation over $a$ is now understood, with which the constrained path integral can be written as a Fourier transform, just like in \cref{eq:defZeta} of the main text,
	\begin{equation}
	    \Zphi=\lim_{\varepsilon\to0}\int_{-\infty}^\infty \Nmprod[a] \frac{\dr[\DDR_a]}{2\uppi}\,\Zeta\,\eu^{\iu\DDR_a\phi_a-\DDR_a^2\varepsilon/(2V)}\,,
	\end{equation}
	compatible with the superanalysis formula~\cite{DeWitt:2012mdz}.

\bibliography{biblio}

\begin{thebibliography}{43}%
\makeatletter
\providecommand \@ifxundefined [1]{%
 \@ifx{#1\undefined}
}%
\providecommand \@ifnum [1]{%
 \ifnum #1\expandafter \@firstoftwo
 \else \expandafter \@secondoftwo
 \fi
}%
\providecommand \@ifx [1]{%
 \ifx #1\expandafter \@firstoftwo
 \else \expandafter \@secondoftwo
 \fi
}%
\providecommand \natexlab [1]{#1}%
\providecommand \enquote  [1]{``#1''}%
\providecommand \bibnamefont  [1]{#1}%
\providecommand \bibfnamefont [1]{#1}%
\providecommand \citenamefont [1]{#1}%
\providecommand \href@noop [0]{\@secondoftwo}%
\providecommand \href [0]{\begingroup \@sanitize@url \@href}%
\providecommand \@href[1]{\@@startlink{#1}\@@href}%
\providecommand \@@href[1]{\endgroup#1\@@endlink}%
\providecommand \@sanitize@url [0]{\catcode `\\12\catcode `\$12\catcode
  `\&12\catcode `\#12\catcode `\^12\catcode `\_12\catcode `\%12\relax}%
\providecommand \@@startlink[1]{}%
\providecommand \@@endlink[0]{}%
\providecommand \url  [0]{\begingroup\@sanitize@url \@url }%
\providecommand \@url [1]{\endgroup\@href {#1}{\urlprefix }}%
\providecommand \urlprefix  [0]{URL }%
\providecommand \Eprint [0]{\href }%
\providecommand \doibase [0]{https://doi.org/}%
\providecommand \selectlanguage [0]{\@gobble}%
\providecommand \bibinfo  [0]{\@secondoftwo}%
\providecommand \bibfield  [0]{\@secondoftwo}%
\providecommand \translation [1]{[#1]}%
\providecommand \BibitemOpen [0]{}%
\providecommand \bibitemStop [0]{}%
\providecommand \bibitemNoStop [0]{.\EOS\space}%
\providecommand \EOS [0]{\spacefactor3000\relax}%
\providecommand \BibitemShut  [1]{\csname bibitem#1\endcsname}%
\let\auto@bib@innerbib\@empty
\bibitem [{\citenamefont {Brauner}(2010)}]{Brauner:2010wm}%
  \BibitemOpen
  \bibfield  {author} {\bibinfo {author} {\bibfnamefont {T.}~\bibnamefont
  {Brauner}},\ }\bibfield  {title} {\bibinfo {title} {{Spontaneous Symmetry
  Breaking and Nambu-Goldstone Bosons in Quantum Many-Body Systems}},\ }\href
  {https://doi.org/10.3390/sym2020609} {\bibfield  {journal} {\bibinfo
  {journal} {Symmetry}\ }\textbf {\bibinfo {volume} {2}},\ \bibinfo {pages}
  {609} (\bibinfo {year} {2010})},\ \Eprint {https://arxiv.org/abs/1001.5212}
  {arXiv:1001.5212 [hep-th]} \BibitemShut {NoStop}%
\bibitem [{\citenamefont {Maas}(2019)}]{Maas:2017wzi}%
  \BibitemOpen
  \bibfield  {author} {\bibinfo {author} {\bibfnamefont {A.}~\bibnamefont
  {Maas}},\ }\bibfield  {title} {\bibinfo {title} {{Brout-Englert-Higgs
  physics: From foundations to phenomenology}},\ }\href
  {https://doi.org/10.1016/j.ppnp.2019.02.003} {\bibfield  {journal} {\bibinfo
  {journal} {Prog. Part. Nucl. Phys.}\ }\textbf {\bibinfo {volume} {106}},\
  \bibinfo {pages} {132} (\bibinfo {year} {2019})},\ \Eprint
  {https://arxiv.org/abs/1712.04721} {arXiv:1712.04721 [hep-ph]} \BibitemShut
  {NoStop}%
\bibitem [{\citenamefont {Aarts}\ \emph {et~al.}(2023)\citenamefont {Aarts}
  \emph {et~al.}}]{Aarts:2023vsf}%
  \BibitemOpen
  \bibfield  {author} {\bibinfo {author} {\bibfnamefont {G.}~\bibnamefont
  {Aarts}} \emph {et~al.},\ }\bibfield  {title} {\bibinfo {title} {{Phase
  Transitions in Particle Physics}: {Results and Perspectives from Lattice
  Quantum Chromo-Dynamics}},\ }\href
  {https://doi.org/10.1016/j.ppnp.2023.104070} {\bibfield  {journal} {\bibinfo
  {journal} {Prog. Part. Nucl. Phys.}\ }\textbf {\bibinfo {volume} {133}},\
  \bibinfo {pages} {104070} (\bibinfo {year} {2023})},\ \Eprint
  {https://arxiv.org/abs/2301.04382} {arXiv:2301.04382 [hep-lat]} \BibitemShut
  {NoStop}%
\bibitem [{\citenamefont {Hatta}\ and\ \citenamefont
  {Ikeda}(2003)}]{Hatta:2002sj}%
  \BibitemOpen
  \bibfield  {author} {\bibinfo {author} {\bibfnamefont {Y.}~\bibnamefont
  {Hatta}}\ and\ \bibinfo {author} {\bibfnamefont {T.}~\bibnamefont {Ikeda}},\
  }\bibfield  {title} {\bibinfo {title} {{Universality, the QCD critical /
  tricritical point and the quark number susceptibility}},\ }\href
  {https://doi.org/10.1103/PhysRevD.67.014028} {\bibfield  {journal} {\bibinfo
  {journal} {Phys. Rev. D}\ }\textbf {\bibinfo {volume} {67}},\ \bibinfo
  {pages} {014028} (\bibinfo {year} {2003})},\ \Eprint
  {https://arxiv.org/abs/hep-ph/0210284} {arXiv:hep-ph/0210284} \BibitemShut
  {NoStop}%
\bibitem [{\citenamefont {Du}\ \emph {et~al.}(2024)\citenamefont {Du},
  \citenamefont {Sorensen},\ and\ \citenamefont {Stephanov}}]{Du:2024wjm}%
  \BibitemOpen
  \bibfield  {author} {\bibinfo {author} {\bibfnamefont {L.}~\bibnamefont
  {Du}}, \bibinfo {author} {\bibfnamefont {A.}~\bibnamefont {Sorensen}},\ and\
  \bibinfo {author} {\bibfnamefont {M.}~\bibnamefont {Stephanov}},\ }\bibfield
  {title} {\bibinfo {title} {{The QCD phase diagram and Beam Energy Scan
  physics: a theory overview}}\ }(\bibinfo {year} {2024})\ \Eprint
  {https://arxiv.org/abs/2402.10183} {arXiv:2402.10183 [nucl-th]} \BibitemShut
  {NoStop}%
\bibitem [{\citenamefont {Fukuda}\ and\ \citenamefont
  {Kyriakopoulos}()}]{Fukuda:1974ey}%
  \BibitemOpen
  \bibfield  {author} {\bibinfo {author} {\bibfnamefont {R.}~\bibnamefont
  {Fukuda}}\ and\ \bibinfo {author} {\bibfnamefont {E.}~\bibnamefont
  {Kyriakopoulos}},\ }\bibfield  {title} {\bibinfo {title} {Derivation of the
  effective potential},\ }\href {https://doi.org/10.1016/0550-3213(75)90014-0}
  {\bibfield  {journal} {\bibinfo  {journal} {Nucl. Phys. B}\ }\textbf
  {\bibinfo {volume} {85}},\ \bibinfo {pages} {354}}\BibitemShut {NoStop}%
\bibitem [{\citenamefont {O'Raifeartaigh}\ \emph {et~al.}(1986)\citenamefont
  {O'Raifeartaigh}, \citenamefont {Wipf},\ and\ \citenamefont
  {Yoneyama}}]{ORaifeartaigh:1986axd}%
  \BibitemOpen
  \bibfield  {author} {\bibinfo {author} {\bibfnamefont {L.}~\bibnamefont
  {O'Raifeartaigh}}, \bibinfo {author} {\bibfnamefont {A.}~\bibnamefont
  {Wipf}},\ and\ \bibinfo {author} {\bibfnamefont {H.}~\bibnamefont
  {Yoneyama}},\ }\bibfield  {title} {\bibinfo {title} {{The Constraint
  Effective Potential}},\ }\href
  {https://doi.org/10.1016/S0550-3213(86)80031-1} {\bibfield  {journal}
  {\bibinfo  {journal} {Nucl. Phys. B}\ }\textbf {\bibinfo {volume} {271}},\
  \bibinfo {pages} {653} (\bibinfo {year} {1986})}\BibitemShut {NoStop}%
\bibitem [{\citenamefont {Endr\H{o}di}\ \emph {et~al.}(2021)\citenamefont
  {Endr\H{o}di}, \citenamefont {Kov\'acs},\ and\ \citenamefont
  {Mark\'o}}]{Endrodi:2021kur}%
  \BibitemOpen
  \bibfield  {author} {\bibinfo {author} {\bibfnamefont {G.}~\bibnamefont
  {Endr\H{o}di}}, \bibinfo {author} {\bibfnamefont {T.~G.}\ \bibnamefont
  {Kov\'acs}},\ and\ \bibinfo {author} {\bibfnamefont {G.}~\bibnamefont
  {Mark\'o}},\ }\bibfield  {title} {\bibinfo {title} {{Spontaneous Symmetry
  Breaking via Inhomogeneities and the Differential Surface Tension}},\ }\href
  {https://doi.org/10.1103/PhysRevLett.127.232002} {\bibfield  {journal}
  {\bibinfo  {journal} {Phys. Rev. Lett.}\ }\textbf {\bibinfo {volume} {127}},\
  \bibinfo {pages} {232002} (\bibinfo {year} {2021})},\ \Eprint
  {https://arxiv.org/abs/2109.03668} {arXiv:2109.03668 [hep-lat]} \BibitemShut
  {NoStop}%
\bibitem [{\citenamefont {Endr\H{o}di}\ \emph {et~al.}(2024)\citenamefont
  {Endr\H{o}di}, \citenamefont {Kov\'acs}, \citenamefont {Mark\'o},\ and\
  \citenamefont {Pannullo}}]{Endrodi:2023est}%
  \BibitemOpen
  \bibfield  {author} {\bibinfo {author} {\bibfnamefont {G.}~\bibnamefont
  {Endr\H{o}di}}, \bibinfo {author} {\bibfnamefont {T.~G.}\ \bibnamefont
  {Kov\'acs}}, \bibinfo {author} {\bibfnamefont {G.}~\bibnamefont {Mark\'o}},\
  and\ \bibinfo {author} {\bibfnamefont {L.}~\bibnamefont {Pannullo}},\
  }\bibfield  {title} {\bibinfo {title} {{Flattening of the quantum effective
  potential in fermionic theories}},\ }\href
  {https://doi.org/10.22323/1.453.0379} {\bibfield  {journal} {\bibinfo
  {journal} {PoS}\ }\textbf {\bibinfo {volume} {LATTICE2023}},\ \bibinfo
  {pages} {379} (\bibinfo {year} {2024})},\ \Eprint
  {https://arxiv.org/abs/2312.01960} {arXiv:2312.01960 [hep-lat]} \BibitemShut
  {NoStop}%
\bibitem [{\citenamefont {Azcoiti}\ \emph {et~al.}(1995)\citenamefont
  {Azcoiti}, \citenamefont {Laliena},\ and\ \citenamefont
  {Luo}}]{Azcoiti:1995dq}%
  \BibitemOpen
  \bibfield  {author} {\bibinfo {author} {\bibfnamefont {V.}~\bibnamefont
  {Azcoiti}}, \bibinfo {author} {\bibfnamefont {V.}~\bibnamefont {Laliena}},\
  and\ \bibinfo {author} {\bibfnamefont {X.-Q.}\ \bibnamefont {Luo}},\
  }\bibfield  {title} {\bibinfo {title} {{Spontaneous symmetry breaking in
  fermion - gauge systems: A Nonstandard approach}},\ }\href
  {https://doi.org/10.1016/0370-2693(95)00602-H} {\bibfield  {journal}
  {\bibinfo  {journal} {Phys. Lett. B}\ }\textbf {\bibinfo {volume} {354}},\
  \bibinfo {pages} {111} (\bibinfo {year} {1995})},\ \Eprint
  {https://arxiv.org/abs/hep-th/9509091} {arXiv:hep-th/9509091} \BibitemShut
  {NoStop}%
\bibitem [{\citenamefont {Luo}(2004)}]{Luo:2004wx}%
  \BibitemOpen
  \bibfield  {author} {\bibinfo {author} {\bibfnamefont {X.-Q.}\ \bibnamefont
  {Luo}},\ }\bibfield  {title} {\bibinfo {title} {{Chiral condensate of lattice
  QCD with massless quarks from the probability distribution function
  method}},\ }\href {https://doi.org/10.1103/PhysRevD.69.076012} {\bibfield
  {journal} {\bibinfo  {journal} {Phys. Rev. D}\ }\textbf {\bibinfo {volume}
  {69}},\ \bibinfo {pages} {076012} (\bibinfo {year} {2004})},\ \Eprint
  {https://arxiv.org/abs/hep-lat/0409032} {arXiv:hep-lat/0409032} \BibitemShut
  {NoStop}%
\bibitem [{\citenamefont {Brodsky}\ and\ \citenamefont
  {Shrock}(2011)}]{Brodsky:2009zd}%
  \BibitemOpen
  \bibfield  {author} {\bibinfo {author} {\bibfnamefont {S.~J.}\ \bibnamefont
  {Brodsky}}\ and\ \bibinfo {author} {\bibfnamefont {R.}~\bibnamefont
  {Shrock}},\ }\bibfield  {title} {\bibinfo {title} {{Condensates in Quantum
  Chromodynamics and the Cosmological Constant}},\ }\href
  {https://doi.org/10.1073/pnas.1010113107} {\bibfield  {journal} {\bibinfo
  {journal} {Proc. Nat. Acad. Sci.}\ }\textbf {\bibinfo {volume} {108}},\
  \bibinfo {pages} {45} (\bibinfo {year} {2011})},\ \Eprint
  {https://arxiv.org/abs/0905.1151} {arXiv:0905.1151 [hep-th]} \BibitemShut
  {NoStop}%
\bibitem [{\citenamefont {Brodsky}\ \emph {et~al.}(2010)\citenamefont
  {Brodsky}, \citenamefont {Roberts}, \citenamefont {Shrock},\ and\
  \citenamefont {Tandy}}]{Brodsky:2010xf}%
  \BibitemOpen
  \bibfield  {author} {\bibinfo {author} {\bibfnamefont {S.~J.}\ \bibnamefont
  {Brodsky}}, \bibinfo {author} {\bibfnamefont {C.~D.}\ \bibnamefont
  {Roberts}}, \bibinfo {author} {\bibfnamefont {R.}~\bibnamefont {Shrock}},\
  and\ \bibinfo {author} {\bibfnamefont {P.~C.}\ \bibnamefont {Tandy}},\
  }\bibfield  {title} {\bibinfo {title} {{Essence of the vacuum quark
  condensate}},\ }\href {https://doi.org/10.1103/PhysRevC.82.022201} {\bibfield
   {journal} {\bibinfo  {journal} {Phys. Rev. C}\ }\textbf {\bibinfo {volume}
  {82}},\ \bibinfo {pages} {022201} (\bibinfo {year} {2010})},\ \Eprint
  {https://arxiv.org/abs/1005.4610} {arXiv:1005.4610 [nucl-th]} \BibitemShut
  {NoStop}%
\bibitem [{\citenamefont {Brodsky}\ \emph {et~al.}(2012)\citenamefont
  {Brodsky}, \citenamefont {Roberts}, \citenamefont {Shrock},\ and\
  \citenamefont {Tandy}}]{Brodsky:2012ku}%
  \BibitemOpen
  \bibfield  {author} {\bibinfo {author} {\bibfnamefont {S.~J.}\ \bibnamefont
  {Brodsky}}, \bibinfo {author} {\bibfnamefont {C.~D.}\ \bibnamefont
  {Roberts}}, \bibinfo {author} {\bibfnamefont {R.}~\bibnamefont {Shrock}},\
  and\ \bibinfo {author} {\bibfnamefont {P.~C.}\ \bibnamefont {Tandy}},\
  }\bibfield  {title} {\bibinfo {title} {{Confinement contains condensates}},\
  }\href {https://doi.org/10.1103/PhysRevC.85.065202} {\bibfield  {journal}
  {\bibinfo  {journal} {Phys. Rev. C}\ }\textbf {\bibinfo {volume} {85}},\
  \bibinfo {pages} {065202} (\bibinfo {year} {2012})},\ \Eprint
  {https://arxiv.org/abs/1202.2376} {arXiv:1202.2376 [nucl-th]} \BibitemShut
  {NoStop}%
\bibitem [{\citenamefont {Sadzikowski}\ and\ \citenamefont
  {Broniowski}(2000)}]{Sadzikowski:2000ap}%
  \BibitemOpen
  \bibfield  {author} {\bibinfo {author} {\bibfnamefont {M.}~\bibnamefont
  {Sadzikowski}}\ and\ \bibinfo {author} {\bibfnamefont {W.}~\bibnamefont
  {Broniowski}},\ }\bibfield  {title} {\bibinfo {title} {Non-uniform chiral
  phase in effective chiral quark models},\ }\href
  {https://doi.org/10.1016/S0370-2693(00)00830-3} {\bibfield  {journal}
  {\bibinfo  {journal} {Phys. Lett. B}\ }\textbf {\bibinfo {volume} {488}},\
  \bibinfo {pages} {63} (\bibinfo {year} {2000})},\ \Eprint
  {https://arxiv.org/abs/hep-ph/0003282} {arXiv:hep-ph/0003282} \BibitemShut
  {NoStop}%
\bibitem [{\citenamefont {Thies}\ and\ \citenamefont
  {Urlichs}(2003)}]{Thies:2003kk}%
  \BibitemOpen
  \bibfield  {author} {\bibinfo {author} {\bibfnamefont {M.}~\bibnamefont
  {Thies}}\ and\ \bibinfo {author} {\bibfnamefont {K.}~\bibnamefont
  {Urlichs}},\ }\bibfield  {title} {\bibinfo {title} {Revised phase diagram of
  the {{Gross-Neveu}} model},\ }\href
  {https://doi.org/10.1103/PhysRevD.67.125015} {\bibfield  {journal} {\bibinfo
  {journal} {Phys. Rev. D}\ }\textbf {\bibinfo {volume} {67}},\ \bibinfo
  {pages} {125015} (\bibinfo {year} {2003})},\ \Eprint
  {https://arxiv.org/abs/hep-th/0302092} {arXiv:hep-th/0302092} \BibitemShut
  {NoStop}%
\bibitem [{\citenamefont {Nakano}\ and\ \citenamefont
  {Tatsumi}(2005)}]{Nakano:2004cd}%
  \BibitemOpen
  \bibfield  {author} {\bibinfo {author} {\bibfnamefont {E.}~\bibnamefont
  {Nakano}}\ and\ \bibinfo {author} {\bibfnamefont {T.}~\bibnamefont
  {Tatsumi}},\ }\bibfield  {title} {\bibinfo {title} {Chiral symmetry and
  density wave in quark matter},\ }\href
  {https://doi.org/10.1103/PhysRevD.71.114006} {\bibfield  {journal} {\bibinfo
  {journal} {Phys. Rev. D}\ }\textbf {\bibinfo {volume} {71}},\ \bibinfo
  {pages} {114006} (\bibinfo {year} {2005})},\ \Eprint
  {https://arxiv.org/abs/hep-ph/0411350} {arXiv:hep-ph/0411350} \BibitemShut
  {NoStop}%
\bibitem [{\citenamefont {Nickel}(2009)}]{Nickel:2009wj}%
  \BibitemOpen
  \bibfield  {author} {\bibinfo {author} {\bibfnamefont {D.}~\bibnamefont
  {Nickel}},\ }\bibfield  {title} {\bibinfo {title} {Inhomogeneous phases in
  the {{Nambu-Jona-Lasino}} and quark-meson model},\ }\href
  {https://doi.org/10.1103/PhysRevD.80.074025} {\bibfield  {journal} {\bibinfo
  {journal} {Phys. Rev. D}\ }\textbf {\bibinfo {volume} {80}},\ \bibinfo
  {pages} {074025} (\bibinfo {year} {2009})},\ \Eprint
  {https://arxiv.org/abs/0906.5295} {arXiv:0906.5295} \BibitemShut {NoStop}%
\bibitem [{\citenamefont {Buballa}\ and\ \citenamefont
  {Carignano}(2015)}]{Buballa:2014tba}%
  \BibitemOpen
  \bibfield  {author} {\bibinfo {author} {\bibfnamefont {M.}~\bibnamefont
  {Buballa}}\ and\ \bibinfo {author} {\bibfnamefont {S.}~\bibnamefont
  {Carignano}},\ }\bibfield  {title} {\bibinfo {title} {Inhomogeneous chiral
  condensates},\ }\href {https://doi.org/10.1016/j.ppnp.2014.11.001} {\bibfield
   {journal} {\bibinfo  {journal} {Progress in Particle and Nuclear Physics}\
  }\textbf {\bibinfo {volume} {81}},\ \bibinfo {pages} {39} (\bibinfo {year}
  {2015})},\ \Eprint {https://arxiv.org/abs/1406.1367} {arXiv:1406.1367
  [hep-ph]} \BibitemShut {NoStop}%
\bibitem [{\citenamefont {Brauner}\ and\ \citenamefont
  {Yamamoto}(2017)}]{Brauner:2016pko}%
  \BibitemOpen
  \bibfield  {author} {\bibinfo {author} {\bibfnamefont {T.}~\bibnamefont
  {Brauner}}\ and\ \bibinfo {author} {\bibfnamefont {N.}~\bibnamefont
  {Yamamoto}},\ }\bibfield  {title} {\bibinfo {title} {{Chiral Soliton Lattice
  and Charged Pion Condensation in Strong Magnetic Fields}},\ }\href
  {https://doi.org/10.1007/JHEP04(2017)132} {\bibfield  {journal} {\bibinfo
  {journal} {JHEP}\ }\textbf {\bibinfo {volume} {04}},\ \bibinfo {pages}
  {132}},\ \Eprint {https://arxiv.org/abs/1609.05213} {arXiv:1609.05213
  [hep-ph]} \BibitemShut {NoStop}%
\bibitem [{\citenamefont {Akerlund}\ \emph {et~al.}(2016)\citenamefont
  {Akerlund}, \citenamefont {de~Forcrand},\ and\ \citenamefont
  {Rindlisbacher}}]{Akerlund:2016myr}%
  \BibitemOpen
  \bibfield  {author} {\bibinfo {author} {\bibfnamefont {O.}~\bibnamefont
  {Akerlund}}, \bibinfo {author} {\bibfnamefont {P.}~\bibnamefont
  {de~Forcrand}},\ and\ \bibinfo {author} {\bibfnamefont {T.}~\bibnamefont
  {Rindlisbacher}},\ }\bibfield  {title} {\bibinfo {title} {{Oscillating
  propagators in heavy-dense QCD}},\ }\href
  {https://doi.org/10.1007/JHEP10(2016)055} {\bibfield  {journal} {\bibinfo
  {journal} {JHEP}\ }\textbf {\bibinfo {volume} {10}},\ \bibinfo {pages}
  {055}},\ \Eprint {https://arxiv.org/abs/1602.02925} {arXiv:1602.02925
  [hep-lat]} \BibitemShut {NoStop}%
\bibitem [{\citenamefont {Haymaker}\ and\ \citenamefont
  {Perez-Mercader}(1983)}]{Haymaker:1983xk}%
  \BibitemOpen
  \bibfield  {author} {\bibinfo {author} {\bibfnamefont {R.~W.}\ \bibnamefont
  {Haymaker}}\ and\ \bibinfo {author} {\bibfnamefont {J.}~\bibnamefont
  {Perez-Mercader}},\ }\bibfield  {title} {\bibinfo {title} {{Convexity of the
  Effective Potential}},\ }\href {https://doi.org/10.1103/PhysRevD.27.1948}
  {\bibfield  {journal} {\bibinfo  {journal} {Phys. Rev. D}\ }\textbf {\bibinfo
  {volume} {27}},\ \bibinfo {pages} {1948} (\bibinfo {year}
  {1983})}\BibitemShut {NoStop}%
\bibitem [{\citenamefont {Alexandre}\ and\ \citenamefont
  {Tsapalis}(2013)}]{Alexandre:2012ht}%
  \BibitemOpen
  \bibfield  {author} {\bibinfo {author} {\bibfnamefont {J.}~\bibnamefont
  {Alexandre}}\ and\ \bibinfo {author} {\bibfnamefont {A.}~\bibnamefont
  {Tsapalis}},\ }\bibfield  {title} {\bibinfo {title} {{Maxwell Construction
  for Scalar Field Theories with Spontaneous Symmetry Breaking}},\ }\href
  {https://doi.org/10.1103/PhysRevD.87.025028} {\bibfield  {journal} {\bibinfo
  {journal} {Phys. Rev. D}\ }\textbf {\bibinfo {volume} {87}},\ \bibinfo
  {pages} {025028} (\bibinfo {year} {2013})},\ \Eprint
  {https://arxiv.org/abs/1211.0921} {arXiv:1211.0921 [hep-th]} \BibitemShut
  {NoStop}%
\bibitem [{\citenamefont {Weinberg}\ and\ \citenamefont
  {Wu}(1987)}]{Weinberg:1987vp}%
  \BibitemOpen
  \bibfield  {author} {\bibinfo {author} {\bibfnamefont {E.~J.}\ \bibnamefont
  {Weinberg}}\ and\ \bibinfo {author} {\bibfnamefont {A.-q.}\ \bibnamefont
  {Wu}},\ }\bibfield  {title} {\bibinfo {title} {{Understanding complex
  perturbative effective potentials}},\ }\href
  {https://doi.org/10.1103/PhysRevD.36.2474} {\bibfield  {journal} {\bibinfo
  {journal} {Phys. Rev. D}\ }\textbf {\bibinfo {volume} {36}},\ \bibinfo
  {pages} {2474} (\bibinfo {year} {1987})}\BibitemShut {NoStop}%
\bibitem [{\citenamefont {DeWitt}(2012)}]{DeWitt:2012mdz}%
  \BibitemOpen
  \bibfield  {author} {\bibinfo {author} {\bibfnamefont {B.~S.}\ \bibnamefont
  {DeWitt}},\ }\href {https://doi.org/10.1017/CBO9780511564000} {\emph
  {\bibinfo {title} {{Supermanifolds}}}},\ Cambridge Monographs on Mathematical
  Physics\ (\bibinfo  {publisher} {Cambridge Univ. Press},\ \bibinfo {address}
  {Cambridge, UK},\ \bibinfo {year} {2012})\BibitemShut {NoStop}%
\bibitem [{\citenamefont {Bender}\ and\ \citenamefont
  {Orszag}(1999)}]{Bender:1999box}%
  \BibitemOpen
  \bibfield  {author} {\bibinfo {author} {\bibfnamefont {C.~M.}\ \bibnamefont
  {Bender}}\ and\ \bibinfo {author} {\bibfnamefont {S.~A.}\ \bibnamefont
  {Orszag}},\ }\href {https://doi.org/10.1007/978-1-4757-3069-2} {\emph
  {\bibinfo {title} {Advanced {{Mathematical Methods}} for {{Scientists}} and
  {{Engineers I}}}}}\ (\bibinfo  {publisher} {Springer},\ \bibinfo {year}
  {1999})\BibitemShut {NoStop}%
\bibitem [{\citenamefont {Langfeld}(2017)}]{Langfeld:2016kty}%
  \BibitemOpen
  \bibfield  {author} {\bibinfo {author} {\bibfnamefont {K.}~\bibnamefont
  {Langfeld}},\ }\bibfield  {title} {\bibinfo {title} {{Density-of-states}},\
  }\href {https://doi.org/10.22323/1.256.0010} {\bibfield  {journal} {\bibinfo
  {journal} {PoS}\ }\textbf {\bibinfo {volume} {LATTICE2016}},\ \bibinfo
  {pages} {010} (\bibinfo {year} {2017})},\ \Eprint
  {https://arxiv.org/abs/1610.09856} {arXiv:1610.09856 [hep-lat]} \BibitemShut
  {NoStop}%
\bibitem [{\citenamefont {Lucini}\ \emph {et~al.}()\citenamefont {Lucini},
  \citenamefont {Mason}, \citenamefont {Piai}, \citenamefont {Rinaldi},\ and\
  \citenamefont {Vadacchino}}]{Lucini:2023irm}%
  \BibitemOpen
  \bibfield  {author} {\bibinfo {author} {\bibfnamefont {B.}~\bibnamefont
  {Lucini}}, \bibinfo {author} {\bibfnamefont {D.}~\bibnamefont {Mason}},
  \bibinfo {author} {\bibfnamefont {M.}~\bibnamefont {Piai}}, \bibinfo {author}
  {\bibfnamefont {E.}~\bibnamefont {Rinaldi}},\ and\ \bibinfo {author}
  {\bibfnamefont {D.}~\bibnamefont {Vadacchino}},\ }\bibfield  {title}
  {\bibinfo {title} {First-order phase transitions in {{Yang-Mills}} theories
  and the density of state method},\ }\href
  {https://doi.org/10.1103/PhysRevD.108.074517} {\bibfield  {journal} {\bibinfo
   {journal} {Phys. Rev. D}\ }\textbf {\bibinfo {volume} {108}},\ \bibinfo
  {pages} {074517}}\BibitemShut {NoStop}%
\bibitem [{\citenamefont {Endr\H{o}di}\ \emph {et~al.}(2018)\citenamefont
  {Endr\H{o}di}, \citenamefont {Fodor}, \citenamefont {Katz}, \citenamefont
  {Sexty}, \citenamefont {Szab{\'o}},\ and\ \citenamefont
  {T{\"o}r{\"o}k}}]{Endrodi:2018zda}%
  \BibitemOpen
  \bibfield  {author} {\bibinfo {author} {\bibfnamefont {G.}~\bibnamefont
  {Endr\H{o}di}}, \bibinfo {author} {\bibfnamefont {Z.}~\bibnamefont {Fodor}},
  \bibinfo {author} {\bibfnamefont {S.~D.}\ \bibnamefont {Katz}}, \bibinfo
  {author} {\bibfnamefont {D.}~\bibnamefont {Sexty}}, \bibinfo {author}
  {\bibfnamefont {K.~K.}\ \bibnamefont {Szab{\'o}}},\ and\ \bibinfo {author}
  {\bibfnamefont {C.}~\bibnamefont {T{\"o}r{\"o}k}},\ }\bibfield  {title}
  {\bibinfo {title} {{Applying constrained simulations for low temperature
  lattice QCD at finite baryon chemical potential}},\ }\href
  {https://doi.org/10.1103/PhysRevD.98.074508} {\bibfield  {journal} {\bibinfo
  {journal} {Phys. Rev. D}\ }\textbf {\bibinfo {volume} {98}},\ \bibinfo
  {pages} {074508} (\bibinfo {year} {2018})},\ \Eprint
  {https://arxiv.org/abs/1807.08326} {arXiv:1807.08326 [hep-lat]} \BibitemShut
  {NoStop}%
\bibitem [{\citenamefont {Schnetz}\ \emph {et~al.}()\citenamefont {Schnetz},
  \citenamefont {Thies},\ and\ \citenamefont {Urlichs}}]{Schnetz:2005ih}%
  \BibitemOpen
  \bibfield  {author} {\bibinfo {author} {\bibfnamefont {O.}~\bibnamefont
  {Schnetz}}, \bibinfo {author} {\bibfnamefont {M.}~\bibnamefont {Thies}},\
  and\ \bibinfo {author} {\bibfnamefont {K.}~\bibnamefont {Urlichs}},\
  }\bibfield  {title} {\bibinfo {title} {Full {{Phase Diagram}} of the
  {{Massive Gross-Neveu Model}}},\ }\href
  {https://doi.org/10.1016/j.aop.2005.12.007} {\bibfield  {journal} {\bibinfo
  {journal} {Annals of Physics}\ }\textbf {\bibinfo {volume} {321}},\ \bibinfo
  {pages} {2604}},\ \Eprint {https://arxiv.org/abs/hep-th/0511206}
  {hep-th/0511206} \BibitemShut {NoStop}%
\bibitem [{\citenamefont {Basar}\ \emph {et~al.}()\citenamefont {Basar},
  \citenamefont {Dunne},\ and\ \citenamefont {Thies}}]{Basar:2009fg}%
  \BibitemOpen
  \bibfield  {author} {\bibinfo {author} {\bibfnamefont {G.}~\bibnamefont
  {Basar}}, \bibinfo {author} {\bibfnamefont {G.~V.}\ \bibnamefont {Dunne}},\
  and\ \bibinfo {author} {\bibfnamefont {M.}~\bibnamefont {Thies}},\ }\bibfield
   {title} {\bibinfo {title} {Inhomogeneous {{Condensates}} in the
  {{Thermodynamics}} of the {{Chiral NJL}}\_2 model},\ }\href
  {https://doi.org/10.1103/PhysRevD.79.105012} {\bibfield  {journal} {\bibinfo
  {journal} {Phys. Rev. D}\ }\textbf {\bibinfo {volume} {79}},\ \bibinfo
  {pages} {105012}},\ \Eprint {https://arxiv.org/abs/0903.1868} {0903.1868}
  \BibitemShut {NoStop}%
\bibitem [{\citenamefont {Gross}\ and\ \citenamefont {Neveu}()}]{Gross:1974jv}%
  \BibitemOpen
  \bibfield  {author} {\bibinfo {author} {\bibfnamefont {D.~J.}\ \bibnamefont
  {Gross}}\ and\ \bibinfo {author} {\bibfnamefont {A.}~\bibnamefont {Neveu}},\
  }\bibfield  {title} {\bibinfo {title} {Dynamical symmetry breaking in
  asymptotically free field theories},\ }\href
  {https://doi.org/10.1103/PhysRevD.10.3235} {\bibfield  {journal} {\bibinfo
  {journal} {Phys. Rev. D}\ }\textbf {\bibinfo {volume} {10}},\ \bibinfo
  {pages} {3235}}\BibitemShut {NoStop}%
\bibitem [{\citenamefont {Wolff}()}]{Wolff:1985av}%
  \BibitemOpen
  \bibfield  {author} {\bibinfo {author} {\bibfnamefont {U.}~\bibnamefont
  {Wolff}},\ }\bibfield  {title} {\bibinfo {title} {The phase diagram of the
  infinite {{N Gross-Neveu}} model at finite temperature and chemical
  potential},\ }\href {https://doi.org/10.1016/0370-2693(85)90671-9} {\bibfield
   {journal} {\bibinfo  {journal} {Phys. Lett. B}\ }\textbf {\bibinfo {volume}
  {157}},\ \bibinfo {pages} {303}}\BibitemShut {NoStop}%
\bibitem [{\citenamefont {Hands}\ \emph {et~al.}()\citenamefont {Hands},
  \citenamefont {Mesiti},\ and\ \citenamefont {Worthy}}]{Hands:2020itv}%
  \BibitemOpen
  \bibfield  {author} {\bibinfo {author} {\bibfnamefont {S.}~\bibnamefont
  {Hands}}, \bibinfo {author} {\bibfnamefont {M.}~\bibnamefont {Mesiti}},\ and\
  \bibinfo {author} {\bibfnamefont {J.}~\bibnamefont {Worthy}},\ }\bibfield
  {title} {\bibinfo {title} {Critical behavior in the single flavor
  {{Thirring}} model in 2+{{1D}}},\ }\href
  {https://doi.org/10.1103/PhysRevD.102.094502} {\bibfield  {journal} {\bibinfo
   {journal} {Phys. Rev. D}\ }\textbf {\bibinfo {volume} {102}},\ \bibinfo
  {pages} {094502}}\BibitemShut {NoStop}%
\bibitem [{\citenamefont {Lenz}\ \emph {et~al.}()\citenamefont {Lenz},
  \citenamefont {Wellegehausen},\ and\ \citenamefont {Wipf}}]{Lenz:2019qwu}%
  \BibitemOpen
  \bibfield  {author} {\bibinfo {author} {\bibfnamefont {J.}~\bibnamefont
  {Lenz}}, \bibinfo {author} {\bibfnamefont {B.}~\bibnamefont
  {Wellegehausen}},\ and\ \bibinfo {author} {\bibfnamefont {A.}~\bibnamefont
  {Wipf}},\ }\bibfield  {title} {\bibinfo {title} {Absence of chiral symmetry
  breaking in {{Thirring}} models in 1+2 dimensions},\ }\href
  {https://doi.org/10.1103/PhysRevD.100.054501} {\bibfield  {journal} {\bibinfo
   {journal} {Phys. Rev. D}\ }\textbf {\bibinfo {volume} {100}},\ \bibinfo
  {pages} {054501}},\ \Eprint {https://arxiv.org/abs/1905.00137} {1905.00137}
  \BibitemShut {NoStop}%
\bibitem [{\citenamefont {Pannullo}\ and\ \citenamefont
  {Winstel}()}]{Pannullo:2023one}%
  \BibitemOpen
  \bibfield  {author} {\bibinfo {author} {\bibfnamefont {L.}~\bibnamefont
  {Pannullo}}\ and\ \bibinfo {author} {\bibfnamefont {M.}~\bibnamefont
  {Winstel}},\ }\bibfield  {title} {\bibinfo {title} {Absence of inhomogeneous
  chiral phases in 2+1-dimensional four-fermion and {{Yukawa}} models},\ }\href
  {https://doi.org/10.1103/PhysRevD.108.036011} {\bibfield  {journal} {\bibinfo
   {journal} {Phys. Rev. D}\ }\textbf {\bibinfo {volume} {108}},\ \bibinfo
  {pages} {036011}},\ \Eprint {https://arxiv.org/abs/2305.09444} {2305.09444}
  \BibitemShut {NoStop}%
\bibitem [{\citenamefont {Koenigstein}\ and\ \citenamefont
  {Pannullo}()}]{Koenigstein:2023yzv}%
  \BibitemOpen
  \bibfield  {author} {\bibinfo {author} {\bibfnamefont {A.}~\bibnamefont
  {Koenigstein}}\ and\ \bibinfo {author} {\bibfnamefont {L.}~\bibnamefont
  {Pannullo}},\ }\bibfield  {title} {\bibinfo {title} {Inhomogeneous
  condensation in the {{Gross-Neveu}} model in noninteger spatial dimensions
  \$1\textbackslash leq d {$<$}3\$. {{II}}. {{Nonzero}} temperature and
  chemical potential},\ }\href {https://doi.org/10.1103/PhysRevD.109.056015}
  {\bibfield  {journal} {\bibinfo  {journal} {Phys. Rev. D}\ }\textbf {\bibinfo
  {volume} {109}},\ \bibinfo {pages} {056015}}\BibitemShut {NoStop}%
\bibitem [{\citenamefont {Witten}(1978)}]{Witten:1978qu}%
  \BibitemOpen
  \bibfield  {author} {\bibinfo {author} {\bibfnamefont {E.}~\bibnamefont
  {Witten}},\ }\bibfield  {title} {\bibinfo {title} {Chiral symmetry, the 1/n
  expansion, and the {{SU}}({{N}}) thirring model},\ }\href
  {https://doi.org/10.1016/0550-3213(78)90416-9} {\bibfield  {journal}
  {\bibinfo  {journal} {Nucl. Phys. B}\ }\textbf {\bibinfo {volume} {145}},\
  \bibinfo {pages} {110} (\bibinfo {year} {1978})}\BibitemShut {NoStop}%
\bibitem [{\citenamefont {Ciccone}\ \emph {et~al.}(2024)\citenamefont
  {Ciccone}, \citenamefont {Di~Pietro},\ and\ \citenamefont
  {Serone}}]{Ciccone:2023pdk}%
  \BibitemOpen
  \bibfield  {author} {\bibinfo {author} {\bibfnamefont {R.}~\bibnamefont
  {Ciccone}}, \bibinfo {author} {\bibfnamefont {L.}~\bibnamefont {Di~Pietro}},\
  and\ \bibinfo {author} {\bibfnamefont {M.}~\bibnamefont {Serone}},\
  }\bibfield  {title} {\bibinfo {title} {{Anomalies and persistent order in the
  chiral Gross-Neveu model}},\ }\href {https://doi.org/10.1007/JHEP02(2024)211}
  {\bibfield  {journal} {\bibinfo  {journal} {JHEP}\ }\textbf {\bibinfo
  {volume} {02}},\ \bibinfo {pages} {211}},\ \Eprint
  {https://arxiv.org/abs/2312.13756} {arXiv:2312.13756 [hep-th]} \BibitemShut
  {NoStop}%
\bibitem [{\citenamefont {Lenz}\ \emph {et~al.}(2020)\citenamefont {Lenz},
  \citenamefont {Pannullo}, \citenamefont {Wagner}, \citenamefont
  {Wellegehausen},\ and\ \citenamefont {Wipf}}]{Lenz:2020bxk}%
  \BibitemOpen
  \bibfield  {author} {\bibinfo {author} {\bibfnamefont {J.}~\bibnamefont
  {Lenz}}, \bibinfo {author} {\bibfnamefont {L.}~\bibnamefont {Pannullo}},
  \bibinfo {author} {\bibfnamefont {M.}~\bibnamefont {Wagner}}, \bibinfo
  {author} {\bibfnamefont {B.}~\bibnamefont {Wellegehausen}},\ and\ \bibinfo
  {author} {\bibfnamefont {A.}~\bibnamefont {Wipf}},\ }\bibfield  {title}
  {\bibinfo {title} {{Inhomogeneous phases in the Gross-Neveu model in 1+1
  dimensions at finite number of flavors}},\ }\href
  {https://doi.org/10.1103/PhysRevD.101.094512} {\bibfield  {journal} {\bibinfo
   {journal} {Phys. Rev. D}\ }\textbf {\bibinfo {volume} {101}},\ \bibinfo
  {pages} {094512} (\bibinfo {year} {2020})},\ \Eprint
  {https://arxiv.org/abs/2004.00295} {arXiv:2004.00295 [hep-lat]} \BibitemShut
  {NoStop}%
\bibitem [{\citenamefont {Pannullo}\ \emph {et~al.}(2022)\citenamefont
  {Pannullo}, \citenamefont {Wagner},\ and\ \citenamefont
  {Winstel}}]{Pannullo:2021edr}%
  \BibitemOpen
  \bibfield  {author} {\bibinfo {author} {\bibfnamefont {L.}~\bibnamefont
  {Pannullo}}, \bibinfo {author} {\bibfnamefont {M.}~\bibnamefont {Wagner}},\
  and\ \bibinfo {author} {\bibfnamefont {M.}~\bibnamefont {Winstel}},\
  }\bibfield  {title} {\bibinfo {title} {Inhomogeneous phases in the chirally
  imbalanced 2+1-dimensional {{Gross-Neveu}} model and their absence in the
  continuum limit},\ }\href {https://doi.org/10.3390/sym14020265} {\bibfield
  {journal} {\bibinfo  {journal} {Symmetry}\ }\textbf {\bibinfo {volume}
  {14}},\ \bibinfo {pages} {265} (\bibinfo {year} {2022})},\ \Eprint
  {https://arxiv.org/abs/2112.11183} {2112.11183} \BibitemShut {NoStop}%
\bibitem [{\citenamefont {Winstel}\ and\ \citenamefont
  {Pannullo}(2023)}]{Winstel:2022jkk}%
  \BibitemOpen
  \bibfield  {author} {\bibinfo {author} {\bibfnamefont {M.}~\bibnamefont
  {Winstel}}\ and\ \bibinfo {author} {\bibfnamefont {L.}~\bibnamefont
  {Pannullo}},\ }\bibfield  {title} {\bibinfo {title} {{Stability of
  homogeneous chiral phases against inhomogeneous perturbations in 2+1
  dimensions}},\ }\href {https://doi.org/10.22323/1.430.0195} {\bibfield
  {journal} {\bibinfo  {journal} {PoS}\ }\textbf {\bibinfo {volume}
  {LATTICE2022}},\ \bibinfo {pages} {195} (\bibinfo {year} {2023})},\ \Eprint
  {https://arxiv.org/abs/2211.04414} {arXiv:2211.04414 [hep-ph]} \BibitemShut
  {NoStop}%
\bibitem [{\citenamefont {Pannullo}\ \emph {et~al.}()\citenamefont {Pannullo},
  \citenamefont {Wagner},\ and\ \citenamefont {Winstel}}]{Pannullo:2024sov}%
  \BibitemOpen
  \bibfield  {author} {\bibinfo {author} {\bibfnamefont {L.}~\bibnamefont
  {Pannullo}}, \bibinfo {author} {\bibfnamefont {M.}~\bibnamefont {Wagner}},\
  and\ \bibinfo {author} {\bibfnamefont {M.}~\bibnamefont {Winstel}},\
  }\bibfield  {title} {\bibinfo {title} {Regularization effects in the
  {{Nambu}}-{{Jona-Lasinio}} model: {{Strong}} scheme dependence of
  inhomogeneous phases and persistence of the moat regime},\ }\href
  {https://doi.org/10.1103/PhysRevD.110.076006} {\bibfield  {journal} {\bibinfo
   {journal} {Phys. Rev. D}\ }\textbf {\bibinfo {volume} {110}},\ \bibinfo
  {pages} {076006}}\BibitemShut {NoStop}%
\end{thebibliography}%

\end{document}